\documentclass[fleqn,usenatbib]{mnras}

\usepackage{newtxtext,newtxmath}

\usepackage[T1]{fontenc}
\usepackage{ae,aecompl}
\usepackage{xcolor}
\usepackage{ulem}
\usepackage{soul}

\usepackage{subfig}

\usepackage{graphicx}	
\usepackage{amsmath}	
\usepackage{verbatim}
\usepackage{bm}
\usepackage{multirow}

\newcommand{\change}[1]{{#1}} 

\graphicspath{{./plots/}}

\newcommand{\teff}{$T_{\mathrm{eff}}$ }

\newcommand{\cobold}{\texttt{CO$^5$BOLD}}

\newcommand{\Msun}{$M_{\odot}$}

\newcommand{\gaia}{{\it Gaia}}
\newcommand{\hst}{{\it HST}}

\newcommand{\mDhor}{D_{\rm surf}}
\newcommand{\mDhormax}{D^{\rm max}_{\rm surf}}
\newcommand{\Dhor}{$\mDhor$}
\newcommand{\Dhormax}{$\mDhormax$}

\newcommand{\tsink}{$t_{\rm sink}$}

\newcommand{\MdotDAZ}{\langle\dot{M}_{\rm DAZ, disc}\rangle}
\newcommand{\MDBZ}{\langle M_{\rm DBZ/DZ, disc}\rangle}
\newcommand{\lgMdotDAZ}{\langle\log\dot{M}_{\rm DAZ, disc}\rangle}
\newcommand{\lgMDBZ}{\langle \log M_{\rm DBZ/DZ, disc}\rangle}

\newcommand{\alfven}{Alfv\'{e}n}

\defcitealias{cunningham19}{CU19}

\makeatletter
\DeclareRobustCommand{\I}{%
	\mbox{\check@mathfonts\fontsize\sf@size\z@\selectfont I}%
}
\DeclareRobustCommand{\V}{%
	\mbox{\check@mathfonts\fontsize\sf@size\z@\selectfont V}%
}
\makeatother

\newcommand\Tstrut{\rule{0pt}{3.0ex}}         
\newcommand\Bstrut{\rule[-1.3ex]{0pt}{0pt}}   

\title[White Dwarf Surface Spreading]{Horizontal spreading of planetary debris accreted by white dwarfs}

\author[T. Cunningham et. al]
{Tim Cunningham,$^{1}$\thanks{E-mail: timothy.cunningham@warwick.ac.uk }
Pier-Emmanuel Tremblay,$^{1}$ 
Evan B.~Bauer,$^{2}$ 
Odette Toloza,$^{1}$ 
\newauthor{Elena Cukanovaite,$^{1}$ Detlev Koester,$^{3}$ Jay Farihi,$^{4}$ Bernd Freytag,$^{5}$ Boris T. G\"{a}nsicke,$^{1,6}$}
\newauthor{Hans-G\"{u}nter Ludwig,$^{7}$ and Dimitri Veras$^{1,6}$\thanks{STFC Ernest Rutherford Fellow}}
\\
$^{1}$Department of Physics, University of Warwick, Coventry, CV4 7AL, UK\\
$^{2}$Center for Astrophysics | Harvard \& Smithsonian, 60 Garden St Cambridge, MA 02138, USA \\
$^{3}$Institut f\"{u}r Theoretische Physik und Astrophysik, Universit\"{a}t Kiel, 24098 Kiel, Germany \\
$^{4}$Physics and Astronomy Department, University College London, Gower Street, London, WC1E 6BT, UK \\
$^{5}$Theoretical Astrophysics, Department of Physics and Astronomy, Uppsala University, Box 516, 751 20 Uppsala, Sweden \\
$^{6}$Centre for Exoplanets and Habitability, University of Warwick, Gibbet Hill Road, Coventry CV4 7AL, UK\\
$^{7}$Zentrum f\"{u}r Astronomie der Universit\"{a}t Heidelberg, Landessternwarte, K\"{o}nigstuhl 12, 69117 Heidelberg, Germany
}

\date{Accepted XXX. Received YYY; in original form ZZZ}

\pubyear{2020}

\begin{document}
\label{firstpage}
\pagerange{\pageref{firstpage}--\pageref{lastpage}}
\maketitle

\begin{abstract}
White dwarfs with metal-polluted atmospheres have been studied widely in the context of the accretion of rocky debris from evolved planetary systems. One open question is the geometry of accretion and how material arrives and mixes in the white dwarf surface layers. Using the 3D radiation-hydrodynamics code \cobold, 
we present the first transport coefficients in degenerate star atmospheres which describe the advection-diffusion of a passive scalar across the surface-plane. 
We couple newly derived horizontal diffusion coefficients with previously published vertical diffusion coefficients to provide theoretical constraints on surface spreading of metals in white dwarfs. Our grid of 3D simulations probes the vast majority of the parameter space of convective white dwarfs, with pure-hydrogen atmospheres in the effective temperature range 6000--18\,000\,K and pure-helium atmospheres in the range 12\,000--34\,000\,K. Our results suggest that warm hydrogen-rich atmospheres (DA; $\gtrsim 13\,000$\,K) and helium-rich atmospheres (DB, DBA; $\gtrsim 30\,000$\,K) are unable to efficiently spread the accreted metals across their surface, regardless of the time dependence of accretion. This result may be at odds with the current non-detection of surface abundance variations at white dwarfs with debris discs. For cooler hydrogen- and helium-rich atmospheres, we predict a largely homogeneous distribution of metals across the surface within a vertical diffusion timescale.
This is typically less than 0.1 per cent of disc lifetime estimates, a quantity which is revisited in this paper using the overshoot results.
These results have relevance for studies of the bulk composition of evolved planetary systems and models of accretion disc physics.

\end{abstract}




\section{Introduction}
\label{sec:intro}

It is now accepted that 25--50 per cent of all white dwarfs show clear evidence of metal pollution in their surface layers \citep{zuckerman10,koester14,farihi16}. For most of these objects the unequivocal origin of this material is the evolved planetary system hosted by the white dwarf \citep{jura03,zuckerman07,gaensicke12,vanderburg16,gaensicke19,manser2019,vanderbosch2020,harrison2018,bonsor2020,xu2019}. 
The canonical model is that a circumstellar disc borne of one or many tidally disrupted planetesimals is able to constantly feed planetary material onto the white dwarf surface \citep{jura03}. {
	These metals disappear from the observable layers at rates defined by the sinking timescale \citep{koester09}. In accretion-diffusion equilibrium, the atmospheric abundances remain constant and decrease once accretion stops \citep{dupuis1992}}. Other models suggest that direct impacts of asteroids on star-grazing orbits may provide a small percentage ($\sim$1--2 per cent) of photospheric metal pollution \citep{wyatt14,brown2017}. {In all cases, surviving planets are required to gravitationally scatter asteroids and minor planets onto the central regions \citep{debes02,debes12,veras16}.}

There have been significant theoretical efforts focused on accretion onto white dwarfs. Of particular importance, \citet{wyatt14} provided a theoretical constraint on the size of particles reaching the white dwarf surface{ and  \citet{turner2020} found that the single-body tidal disruption assumption is invalid in over 20 per cent of cases}. Regarding the geometric configuration of accretion, \citet{metzger12} provided a comprehensive discussion regarding models of discs, finding that magnetic fields as weak as 0.1--1\,kG {at the white dwarf surface} can affect the accreted material, potentially directing it toward the poles. {A predicted outcome of this pole-on accretion is the production of X-rays. This is a detection yet to be made for metal-polluted white dwarfs { \citep{farihi2018XRay}}.}

{Accretion in cataclysmic variables (CVs) is a more mature field than that of metal-polluted white dwarfs. 
For weakly magnetic CVs there are models which favour near-spherical deposition \citep{patterson85} for low accretion rates ($\dot{M}<10^{16}$\,g\,s$^{-1}$) and equatorial deposition {\citep{piro}} for high accretion rates ($\dot{M}>10^{16}$\,g\,s$^{-1}$). The regime of metal-polluted white dwarfs, where the accretion rates are many orders of magnitude lower ($\dot{M}\sim 10^{9}$~g\,s$^{-1}$, {see, e.g., \citealt{farihi16}), has however not been widely explored.}}

The sinking timescales of trace elements through white dwarf atmospheres has received, and continues to receive, considerable attention \citep{schatzman45,paquette86a,paquette86b,dupuis1992,koester09,bauer2019,cunningham19,koester2020,heinonen2020}. {White dwarfs develop convective atmospheres and envelopes below 18\,000~K \citep[hydrogen-rich;][]{cunningham19} or 40--60\,000~K \citep[helium-rich;][]{bergeron2011,cukanovaite19}}. In the accretion-diffusion model \citep{dupuis1992,koester09} it is normally assumed that for white dwarfs with a convective atmosphere,  the observed spectroscopic abundance is distributed homogeneously throughout the mixed region, in the vertical and horizontal directions.

Rapid convective mixing in the vertical (or radial) direction is well-justified given {the shallow convection zones restricted to the hydrogen or helium envelopes (depth of 0.1--100\,km) and short turnover timescales in the range $t_{\rm c}$ = 0.1--1000\,s \citep{vangrootel12}}.
On the other hand, the geometric extent of convection zones in the horizontal (or surface-plane, or azimuthal) directions can be many orders of magnitude larger; with a typical white dwarf circumference of $2\pi R_{\rm WD}\sim 50\,000$\,km. Thus the assumption that convection is able to evenly distribute metals over the entire stellar surface is less obvious, {especially if the metals sink vertically before they are able to spread}. A deviation from this assumption could impact the measured photospheric abundances, and hence the accretion rates and accreted masses at these systems. 
A heterogeneous distribution of surface metals may also produce variations in the spectroscopic line strengths and equivalent widths which may be detectable on the rotational period of the white dwarf. 

A theoretical investigation of metal transport across the surface of a white dwarf was made by \citet{montgomery08} who sought to constrain the homogeneity of metals distributed on the surface of the ZZ Ceti G29-38. The authors argued that sinking was faster than lateral spreading and suggested that any observed variability could provide a constraint on the distribution of metals at the surface. However, no variations in metal line strength have been robustly identified in G29-38 \citep{debes08,reach09}, nor any other white dwarf so far \citep{wilson19}.

We revisit the question of sinking versus lateral spreading with 3D radiation-hydrodynamics (RHD) which has recently led to a better understanding of the convective atmospheres of white dwarfs \citep{tremblay13a,kupka18} and enhanced numerical techniques for characterising the advective transport of material \citep{cunningham19}. In particular, we have recently shown that convective overshoot can increase the sinking times in hydrogen-atmosphere white dwarfs by two orders of magnitude. This will directly impact any conclusion about the balance of sinking times and lateral spreading times.

In Section\,\ref{sec:numerics} we describe our numerical approach to the problem. In Section\,\ref{sec:results} we present results from our 3D model atmosphere simulations. Sections\,\ref{sec:discussion-WDpop} and \ref{sec:discussion} focus on the implications of our results on the white dwarf population and finally we provide conclusions in Section\,\ref{sec:conclusions}.

\section{Numerical Setup}
\label{sec:numerics}
Our approach is to model the transport of trace metals across the white dwarf surface as an effective diffusion process.
This requires the local diffusion coefficient describing mixing in the radial and angular directions. With a vertical extent of less than $\sim$10~km, a white dwarf convection zone is typically much less than a per cent of the stellar radius, and thus we can simplify the problem to mixing in the radial and horizontal, or surface-plane, directions. In the next two sections we describe first the derivation of the effective diffusion coefficients and secondly the incorporation of these diffusion coefficients into a model describing the surface spreading.

\subsection{Diffusion coefficients}
\begin{figure}
 \centering
 \subfloat{\includegraphics[width=.75\columnwidth]{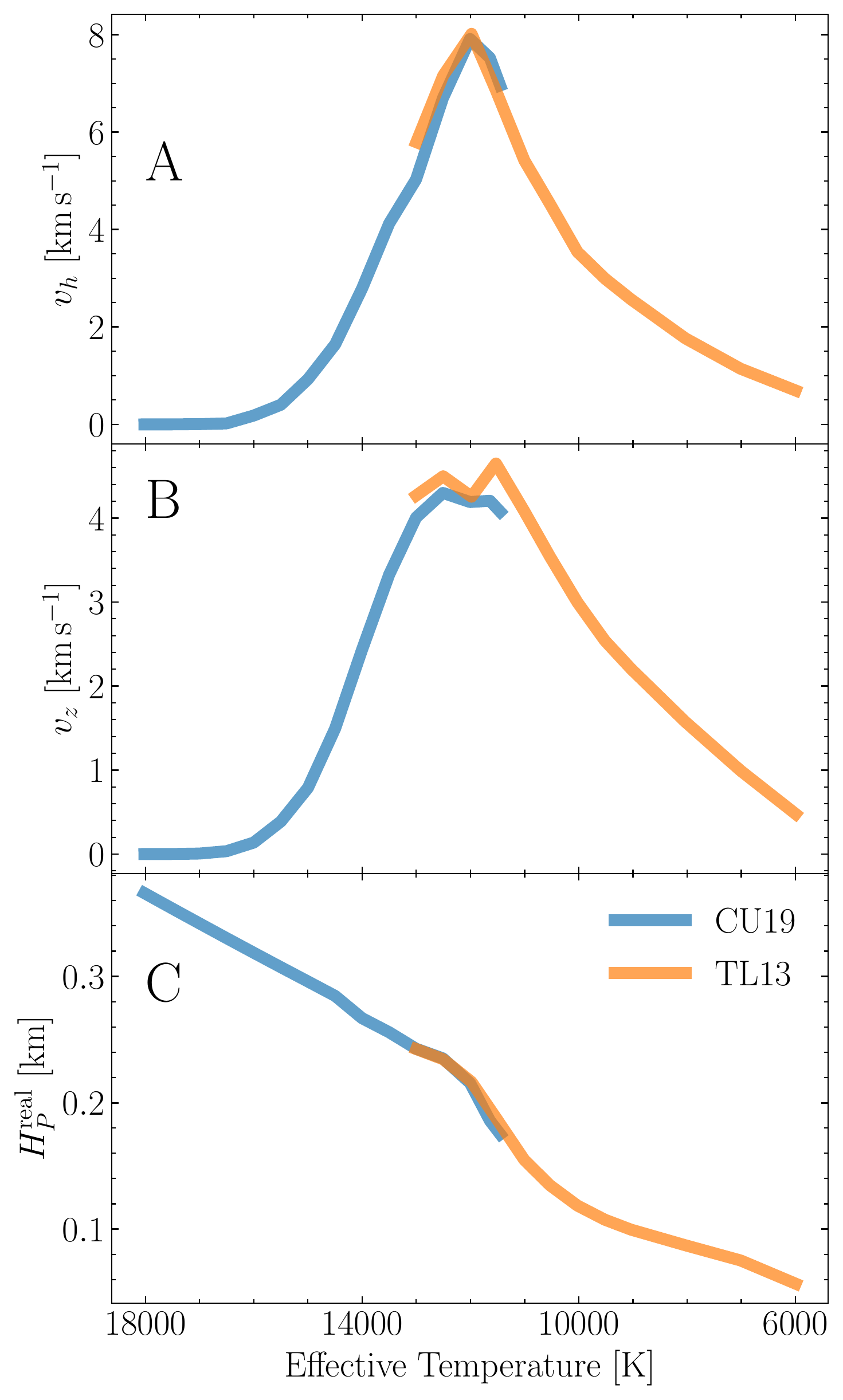}}
 \caption{\textit{Top to bottom:} A) Time-averaged horizontal velocity, $v_h=\sqrt{v_x^2+v_y^2}$, at the photosphere (mean Rosseland optical depth $\langle \tau_{\rm R}\rangle=1$), B) Time-averaged vertical velocity, $v_z$, at the photosphere. C) The actual pressure scale height that corresponds to the geometrical distance between the photosphere and the layer below where the pressure drops by a factor $e$. CU19 and TL13 refer to results from \citetalias{cunningham19} and \citet{tremblay13a} for pure-H atmospheres and are shown in blue  and orange, respectively. The two grids have slightly different numerical setups, with those from \citetalias{cunningham19} (blue) having a greater spatial extent (both deeper and wider) than those of TL13. The resolution is also different between the two grids where the box sizes are $200^3$ and $150^3$ for CU19 and TL13 simulations, respectively. Thus, any difference where the grids overlap does not bare physical significance, but can be attributed to statistical uncertainties between these two grids. We note that for the calculations in this study, we only utilise the parameters $v_h$ (top panel) and $H_{P}^{\mathrm{real}}$ (lower panel).}
 \label{fg:grid_overlap}
\end{figure}

{We present the first 3D RHD \cobold\ \citep{freytag12} simulations designed to directly study the horizontal spreading of trace material in the surface layers of white dwarf stars. 
Our simulations were developed by taking well-relaxed models from two existing grids, adding tracer densities to quantify the horizontal advection-diffusion and running these as a new set of simulations.}
{A grid of deep, RHD simulations for the surface hydrogen convection zones of white dwarfs in the effective temperature range 11\,400--18\,000\,K were presented in \citet[][hereafter CU19]{cunningham19}. Earlier, \citet{tremblay13a} developed a grid of RHD simulations for the surface convection zones of cooler DA white dwarfs with effective temperatures between 6000--13\,500\,K. Here we combine both grids such that the full parameter space of convective DA white dwarfs is probed: from when convective instabilities initially arise at 18\,000\,K down to when the convection zone is sufficiently deep that it couples with the degenerate core. For comparison, we show some key physical properties of both grids in Fig.\,\ref{fg:grid_overlap}. The model grid from \citetalias{cunningham19} is shown in blue and the model grid from \citet{tremblay13a,tremblay13c} is shown in orange. 
The horizontal diffusion coefficients derived from the evolution of a tracer-density are used to confirm that the diffusion coefficient can be computed from mean 3D parameters -- an output of the simulations -- providing a computational advantage. The tracer-density experiments were limited to pure-hydrogen models only, but the parametrisation using mean 3D properties allowed us to exploit the grid of 3D DB (pure-helium) convection zone models from \citet{cukanovaite19} -- spanning effective temperatures of 12\,000--34\,000\,K and surface gravities ($\log g$) of 7.5--9.0 -- to derive horizontal diffusion coefficients for pure-helium atmospheres.}

The \cobold\ simulations presented here, and those previously published, were performed on a Cartesian grid using the \textit{box-in-a-star} setup with periodic side boundary conditions, an open top boundary {and bottom boundary that is either closed or open to convective flows but always open to radiative flux}. The horizontal diffusion coefficients were derived using a tracer density to study the local flows in 3D RHD simulations
and via mean 3D parameters from relaxed simulations. The length scale over which the horizontal diffusion coefficient is calculated is typically an order of magnitude less than the extent of the simulation box and thus the presence of periodic horizontal boundaries is not expected to influence the determination of the horizontal diffusion coefficient. To test this hypothesis, in Section \ref{sec:results-diffusion-coefficients} we perform a convergence test with a simulation domain four times larger in area.

Extending the working of \citetalias{cunningham19}, we trace the effective diffusion in the surface layers of white dwarfs by following the local advective flows in relaxed 3D simulations of surface convection zones.
A simulation is deemed relaxed when the mean properties (i.e., effective temperature, horizontally-averaged velocities) reach a quasi-steady state. This was previously confirmed for all simulations used in this study.
This ensures that the inferred quantities (i.e., horizontal diffusion coefficient) are derived for a simulation which best represents the physical characteristics of a white dwarf at a given effective temperature and surface gravity.

The horizontal diffusion coefficient is calculated by adding vertical slabs of a passive scalar tracer density throughout the simulation domain. In order to probe diffusion in a given direction, a slab must be oriented such that the surface normal is parallel to said direction, i.e., to probe the diffusion in the $x$-direction, a slab must be placed in the $y-z$ plane. Multiple slabs 
($\approx$24--28)
are initialised at evenly spaced locations in the $x$-direction. This allows to quantify the uncertainty on diffusion in the $x$-direction.
The spatial extent of these ensembles, if undergoing a diffusion process, should evolve proportionally to the square-root of time ($t^{1/2}$), such that the diffusion coefficient in the $x$-dimension, $D_x$, can be calculated as
\begin{equation}
2\log_{10}(\sigma_{x}^{*}(x,t)) = \log_{10}(t) + \log_{10}(2D_x)~,
\label{eq:fwhm-D}
\end{equation} 
where $\sigma_{x}^{*}$ is the density-weighted standard deviation of the tracers defined as 
\begin{equation}
\sigma_{x}^{*} = \left(\frac{n_x}{n_x-1}\sum_{x}\left(\rho_{\mathrm{t}}(x)(x - \langle x^* \rangle)^{2}\right)\middle/ \sum_{x} \rho_{\mathrm{t}}(x)\right)^{1/2}~,
\label{eq:sigma_x}
\end{equation}
where $n_x$ is the number of cells in the $x$ dimension, $\rho_{\mathrm{t}}(x)$ is the tracer density mean-averaged across the $y$ and $z$ dimensions, and $\langle x^* \rangle$ is the density-weighted mean $x$-position of the tracer density given by
\begin{equation}
 \langle x^* \rangle = \left(\sum_{x} x \rho_{\mathrm{t}}(x)\right) \cdot \left(\sum_{x} \rho_{\mathrm{t}}(x)\right)^{-1}~.
\end{equation}

\noindent The initial condition for each independent slab of tracer density added at $x=x_i$ is defined as
\begin{equation}
 \rho_{\mathrm{t}}(x) = \begin{cases}
    10^{5} ~\mathrm{cm^{-3}}, & \text{where $x=x_i$},\\
    10^{-6} ~\mathrm{cm^{-3}}, & \text{where $x \neq x_i$}.
  \end{cases}
\end{equation}
As the tracer density, $\rho_t$, is a passive scalar the actual number densities only bare relevance to the precision of the numerical reconstruction schemes and rounding errors. The initial positions, $x_i$, are chosen to evenly sample the full horizontal extent of the simulation box in the $x$-dimension. This procedure is repeated in the $y$-dimension such that an alternative component to the horizontal diffusion coefficient, $D_y$, is found by placing slabs of a passive scalar density in the $x-z$ plane. As it is not expected that the horizontal diffusion coefficient should vary across the $x$- and $y$-dimensions the final diffusion coefficient is computed as a mean across all slabs, such that
\begin{equation}
    \mDhor = \frac{1}{S_x+S_y}\left(\sum_{i=1}^{S_x} D_{x,i} + \sum_{j=1}^{S_y} D_{y,j} \right)~,
    \label{eq:Dhor-mean}
\end{equation}
where $S_x$ and $S_y$ are the number of slabs added in the $x$- and $y$-dimensions, respectively. The standard deviation around this mean quantity (Eq.\,\ref{eq:Dhor-mean}) is chosen to represent the error attributed to each horizontal diffusion coefficient. The results of these computations are considered in Section\,\ref{sec:results-diffusion-coefficients} and compared to the well-known 1D estimation of the diffusion coefficient. {We note that this procedure ultimately provides a single diffusion coefficient describing transport in the surface plane. This omits any radial dependence of the diffusion coefficient, and precludes any differential rotation effects. }

In 1D mixing-length theory the local approximation of the diffusion coefficient is typically given (e.g., \citealt{zahn91,ventura98}) as
\begin{equation}
    D = \frac{1}{3} l_{\rm d} v_{\rm conv} \,\,\, ,
    \label{eq:Dmlt}
\end{equation}
where $l_{\rm d}$ is a characteristic convective length scale and $v_{\rm conv}$ is the mean convective velocity. Pressure scale height, $H_{\rm P}$, has often been proposed to be a suitable choice of characteristic mixing length.

Here we explore the validity of this approximation by comparing this expression to results from tracer-density evolution in the convection zone of white dwarfs at nine effective temperatures spanning the full range of convective DA white dwarfs: 6000--18\,000\,K.  The vertical diffusion coefficients for 11\,400--18\,000\,K were presented in \citetalias{cunningham19}. In Section\,\ref{sec:results} we present the results of these experiments, providing the first horizontal diffusion coefficients derived from multi-dimensional simulations in white dwarfs. 

\subsection{Simulating surface transport}
\label{sec:1Dmodel}
Here we introduce the method by which we model the global transport of metals across the surface of the white dwarf. We begin by taking the limiting case of a delta function accretion, both in the temporal and spatial domains. Physically, this effectively represents the collision of an asteroid with the surface of the white dwarf. Modelling the surface of the white dwarf as a ring allows us to exploit the well-known diffusion equation:
\begin{equation}
    \dot{u}(r,t) = \nabla \left(D \nabla u(r,t)\right) .
    \label{eq:heat-eqn}
\end{equation}
Here, $u(r,t)$ is the local concentration of some scalar which, in our case, represents the accreted heavy elements, and $D$ is the diffusion coefficient. As a preliminary scenario we consider the situation that metals are not permitted to sink out of the observable layers. This assumption allows for a simple analytic form of position dependent abundance to be written down. We also assume throughout this study that the diffusion coefficient is independent of the position across the surface. In this case, with only surface (horizontal) transport and a delta function accretion the local concentration can be written as

\begin{equation}
    \frac{\partial u}{\partial t} = \left(\mDhor\right) \frac{\partial^2 u}{\partial x^2},
    \label{eq:1d-heat-eqn}
\end{equation}
where $x$ corresponds to the position along the circumference of the white dwarf and \Dhor\ is the diffusion coefficient describing the effectiveness of material transport in the plane of the surface. This diffusion coefficient is assumed to have no depth dependence for the time being. 

\begin{figure*}
 \centering
 \subfloat{\includegraphics[width=1.5\columnwidth]{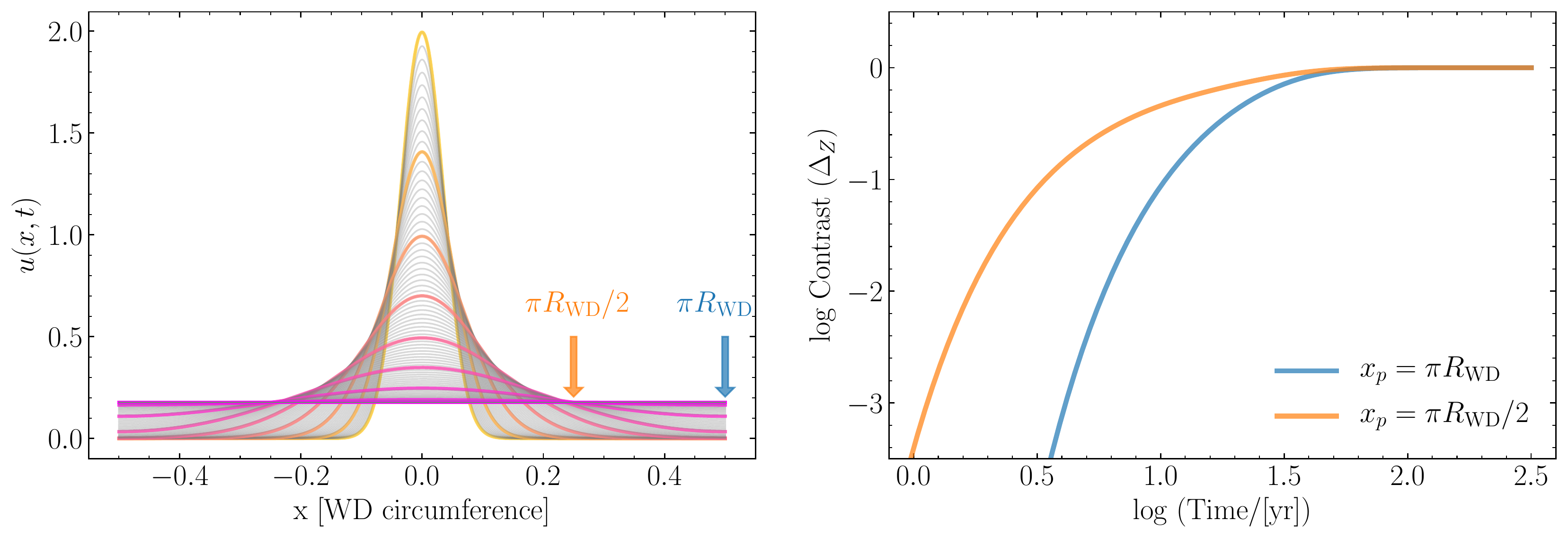}}
 \caption{\textit{Left:} Evolution of a passive scalar, $u$, according to Eq.\,\ref{eq:1d-heat-eqn} with a horizontal diffusion coefficient of $\mDhor = 2\times 10^9$\,cm$^2$s$^{-1}$, characteristic of a white dwarf with $T_{\rm eff}=12\,000$\,K and $\log g=8.0$, and a typical white dwarf radius of $R_{\rm WD}=0.01288$\,$R_{\odot}$. To make clear the evolution, every tenth (logarithmic) time step is indicated with a colour; from yellow to purple. \textit{Right:} Examination of the time-dependent contrast between points of interest along the white dwarf circumference. Here we define contrast ($\Delta z$) according to Eq.\,\ref{eq:contrast} and show the evolution of $\Delta z$ at half and a quarter of the circumference.}
 \label{fg:delta_spread}
\end{figure*}

For spatial domain, $x$, contained in interval $[-L,L]$ and with initial and boundary conditions
\begin{align}
    u(x,0)    &= \delta(x)\\
    u(-L,t)   &= u(L,t)  \\
    u_x(-L,t)   &= u_x(L,t) \, ,
\end{align} 
where $u_x$ denotes the derivative of $u$ with respect to $x$, we can write the analytic solution to the heat equation as 
\begin{equation}
    u(x,t) = \frac{1}{2L} + \frac{1}{L} \sum_{n=1}^{\infty} \cos\left(\frac{n\pi}{L}x\right) e^{-\left(\frac{n\pi}{L}\right)^2Dt}~.
\end{equation}
This study is focused on deriving theoretical constraints on the timescales over which one may or may not expect to observe variability at a single spatially unresolved white dwarf. As a measure of the homogeneity at the white dwarf surface we define the contrast, $\Delta_Z$, between the origin ($x=0$) and an arbitrary point on the surface ($x=x_{i}$) to be
\begin{equation}
    \Delta_Z(x_i, t) = \frac{u(x_i,t)}{u(0,t)} .
    \label{eq:contrast}
\end{equation}
The circumferential positions which will be used to quantify the timescales of interest throughout this study will be the opposite side of the white dwarf ($x=\pi R_{\rm WD}$) and a quarter of a circumference ($x=\pi R_{\rm WD}/2$).
Fig.\,\ref{fg:delta_spread} shows the solution to Eq.\,\eqref{eq:1d-heat-eqn} for the evolution of a delta function scalar on a circular ring of radius $R_{\rm WD}$. This serves as a demonstration of the simple global transport model we employ to estimate the spreading timescales. 

We now go on to present the diffusion coefficients derived from our grid of 3D models. 
We then implement these diffusion coefficients into the simple model outlined previously for simulating surface transport. This is followed by a presentation of the results from a model which accounts for the sinking of material and depth dependence of the diffusion coefficient, though for narrative clarity we introduce that second model as part of the discussion in Section\,\ref{sec:continuous-accretion}.

\section{Results}
\label{sec:results} 

\subsection{Diffusion coefficients}
\label{sec:results-diffusion-coefficients}
\begin{figure}
 \centering
 \subfloat{\includegraphics[width=1\columnwidth]{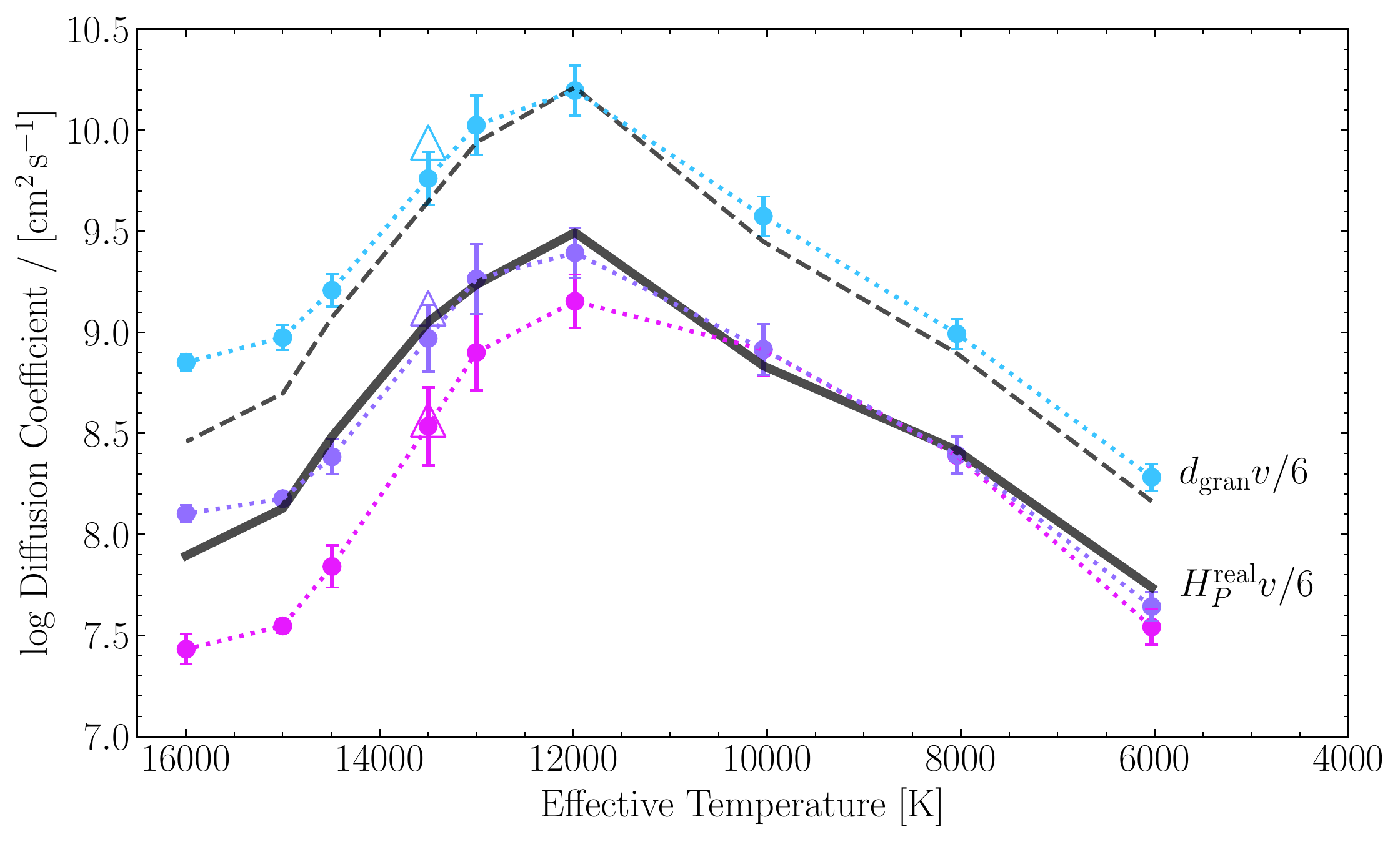}}
 \caption{Effective horizontal diffusion coefficients derived from tracer-density evolution with nine pure-hydrogen \cobold\ simulations (filled circles) between 6000 and 15\,000\,K at $\log g = 8.0$ with grey opacities. The tracer distribution is averaged over all depths of a simulation. Models with $T_{\rm eff}\geqslant 12\,000$\,K have closed bottom boundaries, as opposed to the cooler models with open bottom boundaries. For the closed bottom simulations, the trace density in the stable lower layers serves to reduce the effective diffusion coefficient by averaging over layers outside of the convectively mixed region. To mitigate this effect we employ a clipping procedure at $<$100, $<$10 and $<$0.01 per cent of the initial maximum tracer density in pink (i.e., no clipping), purple and cyan, respectively (see Fig.\,\ref{fg:qucs_t135}). A lower clipping value corresponds to probing increasingly fast and rare local diffusion events within the resolved simulation. The cyan line, representing the fastest diffusing layers or resolved diffusion events, can be considered a robust upper limit on the horizontal diffusion coefficient for passive scalars. The purple line corresponds to our best estimate of a mean diffusion coefficient across the accessible convectively mixed layers. The black solid line shows the analytic approximation of the local diffusion coefficient given by Eq.\,\eqref{eq:Dreal}. The black dashed line shows an alternative analytic prescription for the diffusion coefficient where the length scale is chosen to be the characteristic granule size, rather than the pressure scale height.}
 \label{fg:direct_full}
\end{figure}

\begin{figure}
 \centering
 \subfloat{\includegraphics[width=1.\columnwidth]{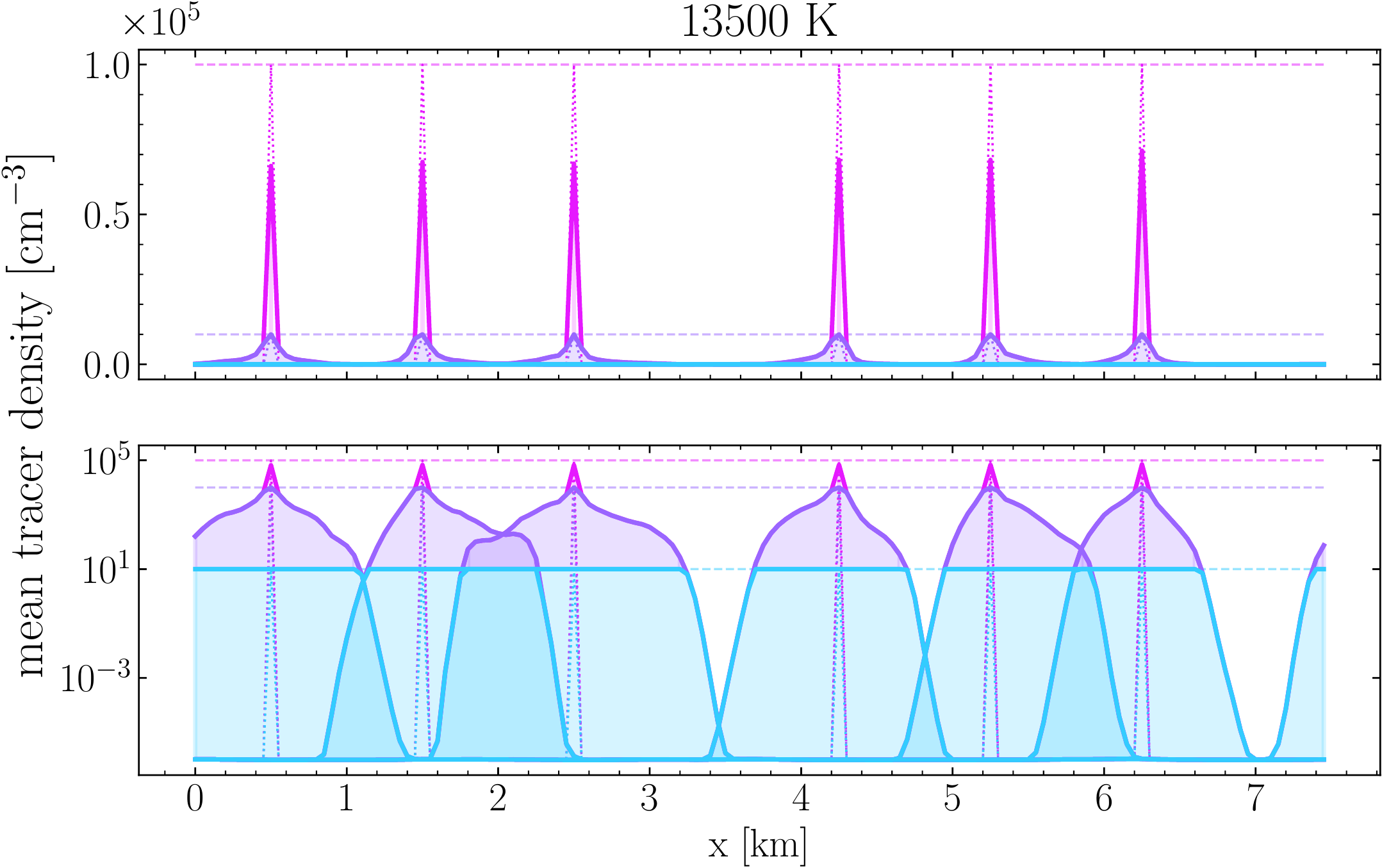}}
 \caption{A selection of mean tracer densities, $\rho_t(x)$, in the $x$ dimension, averaged over the vertical ($z$) and other horizontal ($y$) dimension, for a simulation at 13\,500\,K and $\log g = 8.0$, and in the time interval $t=0$\,s (dotted) and $t=0.1$\,s (solid). The profiles are shown both linearly (top panel) and logarithmically (bottom panel) for clarity. To remove the contribution of the (non-diffusing) stable lower layers from the mean tracer densities, the distributions are clipped at $<$10\% (purple) and $<$0.01\% (cyan) of the initial maximum density as described in the text. These colours correspond to the lines shown in Fig.\,\ref{fg:direct_full}. The boundary conditions in the x-dimension are parallel such that the tracer density ($\rho_{\rm tr}$) satisfies $\rho_{\rm tr}(0)=\rho_{\rm tr}(x_{\rm b})$, for $x_{\rm b}=7.5$\,km. {The apparent gap at $x=3.5$\,km is a consequence of plotting only a subset of the 28 tracer density slabs included in this simulation. This omission is for clarity only, and the horizontal diffusion coefficient of each slab is not expected to vary significantly.}}
 \label{fg:qucs_t135}
\end{figure}

The horizontal diffusion coefficient, $\mDhor$, derived from tracer-density evolution on nine simulations between 6000 and 15\,000\,K, are shown in Fig.\,\ref{fg:direct_full}. The tracer experiments were carried out with slabs of tracer density initially placed in the $y-z$ and $x-z$ planes, allowing to follow the diffusion process in the horizontal $x$ and $y$ directions, respectively. Each slab probes the entire vertical extent of the simulation box and does not allow for an explicit depth dependence of $\mDhor$, instead providing a vertically-averaged horizontal diffusion coefficient. The final horizontal diffusion coefficient (and associated error) is computed as the mean (and standard deviation) of the horizontal diffusion coefficient derived for each slab. 

The pink points in Fig.\,\ref{fg:direct_full} represent the mean horizontal diffusion coefficient throughout the entire box. For the closed bottom simulations ($\geq12\,000$\,K) the mean horizontal diffusion coefficient includes a contribution from the non-mixing layers beneath the convective and overshoot region. We account for this contribution by limiting the maximum amplitude of the tracer density distributions (see Fig.\,\ref{fg:qucs_t135}), effectively removing the vertical layers with no or very little mixing. The purple circles in Fig.\,\ref{fg:direct_full} show the horizontal diffusion coefficient when only accounting for the regions where the tracer density is 10 per cent or lower than the maximum initial value. {The arbitrary value of 10 per cent is used to crudely separate the parts of the tracer slab in the deep, non-diffusing layers from those in the shallow, rapidly-diffusing layers. A more robust, but computationally-demanding, approach would be to follow the trajectories of individual tracer particles, preserving the depth dependence of the horizontal diffusion coefficient. A major limitation with this approach however is the large range of advective time scales spanned by a single simulation box, where the vertical diffusion coefficient was found in \citetalias{cunningham19} to vary by up seven orders of magnitude. This crude clipping procedure is thus adopted to disentangle the contributions of horizontal diffusion at different depths, without doing the detailed path integration. The distributions resulting from a clipping value of 10 percent are shown in purple in Fig.\,\ref{fg:qucs_t135}.} 

{To investigate whether this clipping procedure negatively impacts our derived diffusion coefficients we consider its behaviour on the open-bottom and closed-bottom simulations, respectively. The clipping procedure should not alter the open-bottom simulations because the entire simulation domain is contained within the fast-moving photospheric layers. The closed-bottom simulations are those which possess the deep, slowly-diffusing layers, the contribution of which we seek to remove from the derived horizontal diffusion coefficients. We see in Fig.\,\ref{fg:direct_full} that the open-bottom simulations ($T_{\rm eff}<11\,000$\,K) are barely affected by this clipping value of 10 per cent. On the contrary, the closed-bottom simulations ($T_{\rm eff}>11\,000$\,K) exhibit an increase in the diffusion coefficient of 0.2--0.5 dex.}

By extending this approach it is possible to provide an upper limit on the horizontal diffusion coefficient. We probe the most rapidly diffusing layers by considering only the lowest 0.01 per cent of each tracer density distribution, corresponding to the wings of the tracer densities (see Fig.\,\ref{fg:qucs_t135}).
Given that convective velocities peak just below the photosphere \citep{tremblay15}, it is likely that these are the layers being described by this diffusion coefficient (cyan on Fig.\,\ref{fg:direct_full}). In fact, this case is well represented by
\begin{equation}
    \mDhor = \frac{1}{3}\frac{d_{\rm gran}}{2}\sqrt{v_x^2+v_y^2}~,
    \label{eq:Dgran}
\end{equation}
as shown by a dashed black curve in Fig.\,\ref{fg:direct_full}, where the characteristic length is half the characteristic granulation size, $d_{\rm gran}$, defined from the emergent intensity. The granule size corresponds most directly to the characteristic size of convective flows in the photosphere \citep{tremblay13b}, suggesting the equation characterises the mixing capabilities of the uppermost surface layers. We could not clearly establish the reason for the factor of $\frac{1}{2}$ to the granulation size compared to the 1D diffusion equation of Eq.\,\eqref{eq:Dmlt}, but this could be because granulation size is a 2D representation of a 3D process, or that in the definition of granule size
a single granule is composed of both a downdraft and an updraft.

The photosphere only contains a small amount of mass compared to the full convection zone, hence Eq.\,\eqref{eq:Dgran} is unlikely to be a good representation of the overall horizontal mixing of accreted debris in the convection zone. To represent this process, we use instead, as describe above, the clipped estimate where the tracer density is 10 per cent or lower than the maximum initial value. By construction of our models \citep{tremblay13a}, this corresponds to mixing in surface layers above Rosseland optical depth $\tau_{\rm R} \lesssim 1000$. This case is best described by 
\begin{equation}
    \mDhor = \frac{1}{3}\frac{H_{\rm P}^{\rm real}}{2}\sqrt{v_x^2+v_y^2}~.
    \label{eq:Dreal}
\end{equation}
which is shown by a solid black curve in Fig.\,\ref{fg:direct_full}. The characteristic length $H_{\rm P}^{\rm real}$ is the actual pressure scale height that corresponds to the geometrical distance between the photosphere ($\langle \tau_{\rm R}\rangle=1$) and the layer below where the pressure drops by a factor $e$.
It is not clear whether the factor of $H_{\rm P}^{\rm real}/2$ in Eq.\,\eqref{eq:Dreal}, when compared to Eq.\,\eqref{eq:Dmlt}, has a fundamental physical justification, as here the diffusion process corresponds to an average over a range of convective layers.

For open bottom simulations with \teff $<$ 12\,000\,K, Eq.\,\eqref{eq:Dreal} may still represent an upper limit on the efficiency of the overall diffusion of accreted material across the stellar surface. The reason is that the lower portions of the convection zones, which are not included in open bottom simulations, have lower convective velocities and are possibly less efficient to mix material. While this would not necessarily delay the process of achieving a fully homogeneous surface composition across the stellar disc, the actual absolute metal abundances could vary on longer timescales, as the bottom of the convective zone is still adjusting. A lower limit on horizontal diffusion could possibly rely on using the pressure scale height at the bottom of the convection zone in Eq.\,\eqref{eq:Dmlt}.

We provide the results of the tracer-density evolution experiments in Table\,\ref{tb:D3d} where one can find the mean horizontal diffusion coefficient in the mixed layers (purple circles in Fig.\,\ref{fg:direct_full}) as $\log \mDhor$. We also provide the upper limits (corresponding to the photosphere) on the diffusion coefficient at each effective temperature as $\log \mDhormax$. 

To test the convergence of our results we also performed a direct tracer experiment on a simulation with twice the spatial extent in the x and y dimensions. From \citetalias{cunningham19} we adapted the simulation C1-2 which was identical in setup to simulation C1 from the same study -- with $T_{\mathrm{eff}}=13,500$\,K, $\log g=8.0$ and box size of $150^3$ -- except that the x and y dimensions each extended to 15\,km, rather than 7.5\,km. The results of direct tracer experiments for this simulation are shown in Fig.\,\ref{fg:direct_full} with open triangles. We find the results of this simulation (C1-2) to be in reasonable agreement with the one with a lesser spatial extent (C1). The minimum diffusion coefficient (pink), which includes the slowest diffusing layers, is in excellent agreement, whilst the most physical estimate for the diffusion coefficient ($D_{\rm surf}$; purple), after the removal of the contribution from the overshoot layers, agrees to within a standard deviation.

By way of comparison to previous studies, \citet{montgomery08} derived a horizontal diffusion coefficient of $1.5 \times 10^{10}$\,cm$^2$ s$^{-1}$ for the metal-polluted white dwarf G29-38 using white dwarf evolutionary models and stellar parameters constrained by observations. The derived parameters for this star are $T_{\rm eff}=11\,400$\,K and $\log g = 8.02$ \citep{gentile2019} and using these we can compare the horizontal diffusion coefficient of \citet{montgomery08} to those shown in Fig.\,\ref{fg:direct_full}. We see that it is in reasonable agreement with our predicted maximum value \Dhormax\ (blue) corresponding to the photosphere. However, it is about an order of magnitude larger than our adopted coefficient (purple) considering mixing over the entire convection zone.

\begin{figure}
	\centering
	\subfloat{\includegraphics[width=1\columnwidth]{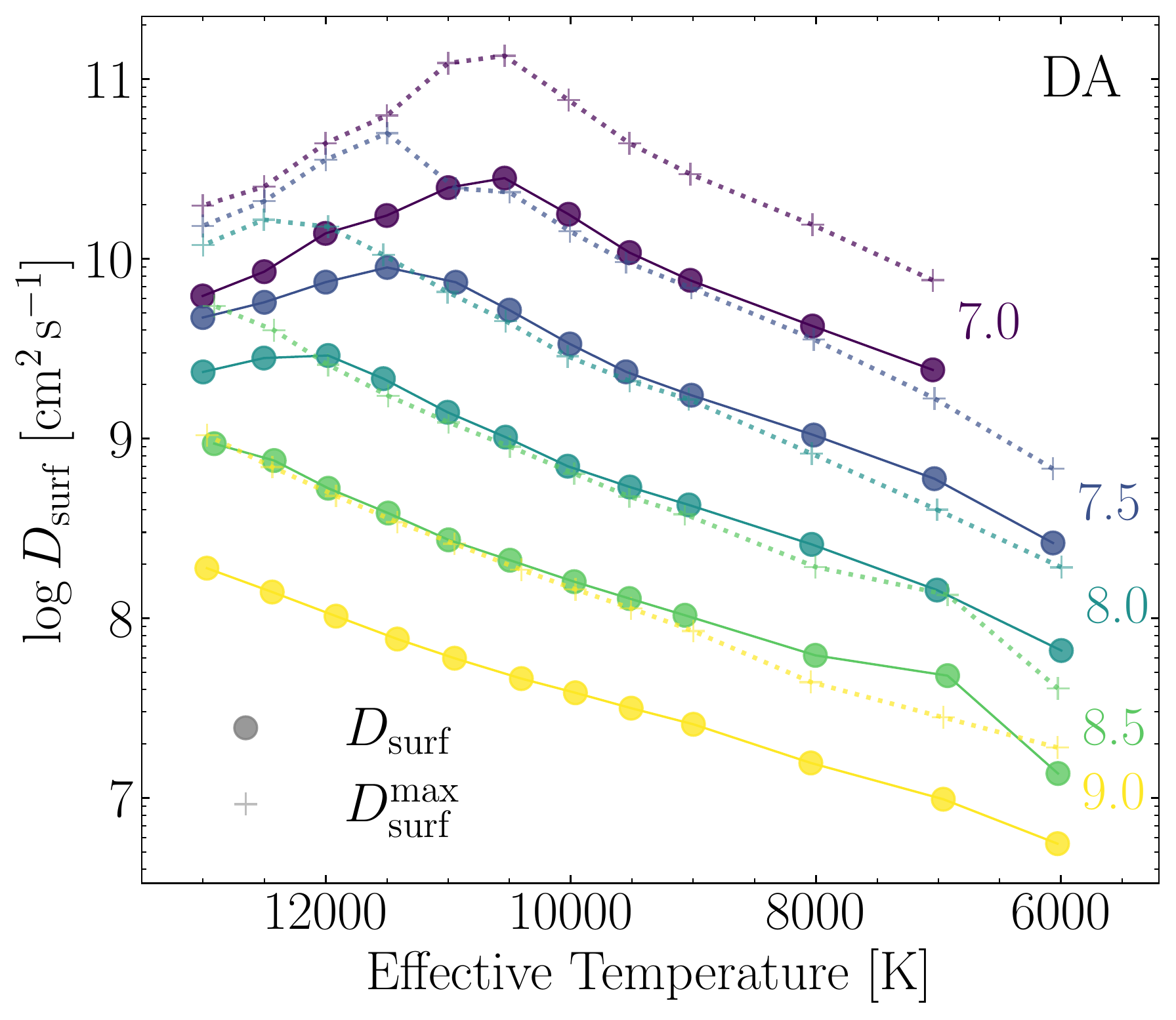}}
	\caption{Horizontal diffusion coefficients for pure-hydrogen white dwarf models defined using mean 3D values derived from the grid of models presented in \citet{tremblay13a}. Diffusion coefficients are derived using Eq.\,\eqref{eq:Dreal} (circles, solid lines) and Eq.\,\eqref{eq:Dgran} (plusses, dotted lines). The latter represents photospheric diffusion which serves as an upper limit on the horizontal diffusion coefficient, \Dhormax. The mean error on the diffusion coefficients derived from direct diffusion experiments (Table\,\ref{tb:D3d}) was found to be $\sigma(\log \mDhor)=0.1$ which we adopt as an estimate for the error on points shown in this figure. The surface gravity ($\log g$) of each line pair is shown in the corresponding colour. All values shown here are displayed in Table\,\ref{tb:Dderived_DA_full}. }
	\label{fg:Dderived}
\end{figure}

\begin{figure}
	\centering
	\subfloat{\includegraphics[width=1.\columnwidth]{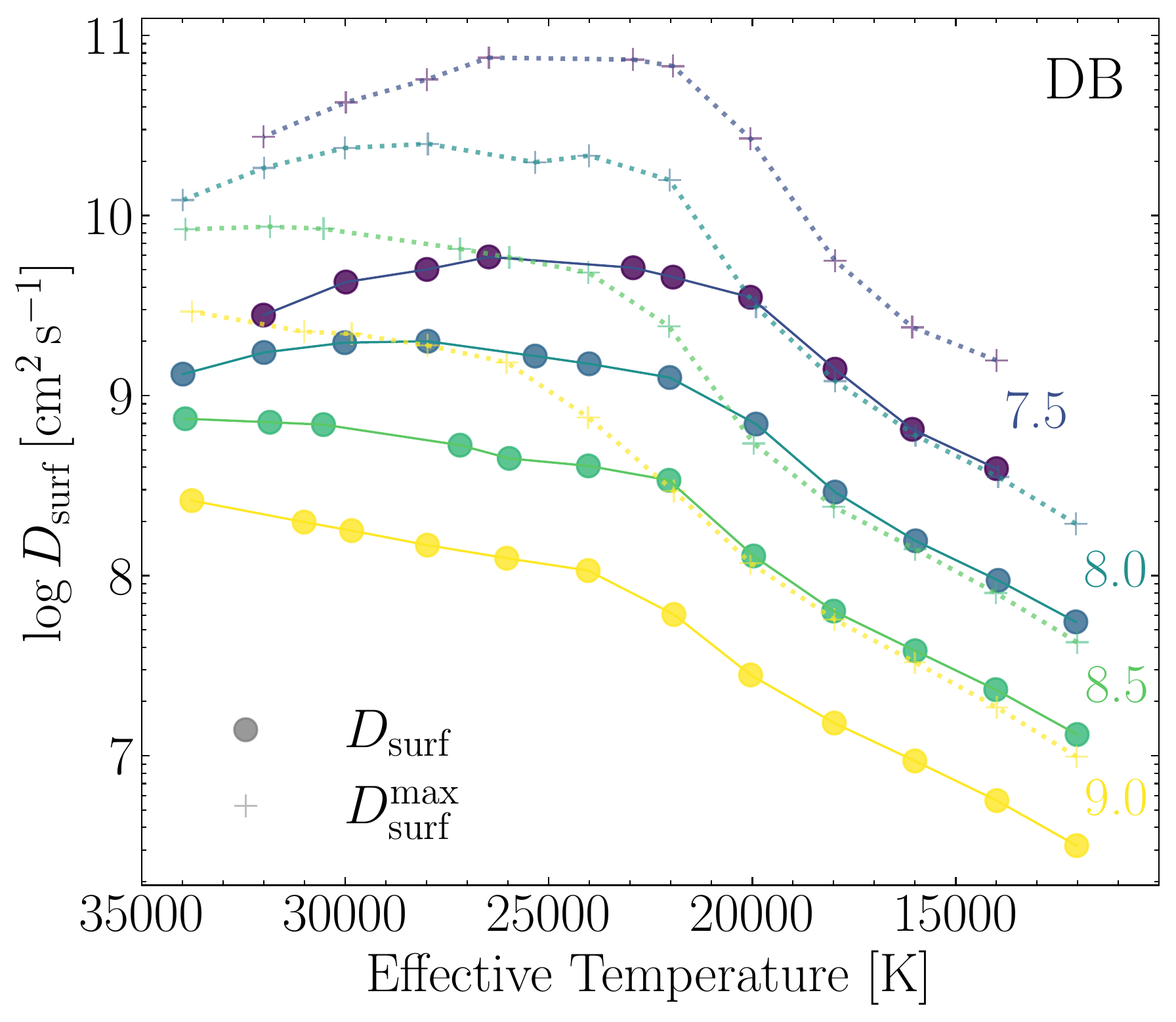}}
	\caption{Similar to Fig.\,\ref{fg:Dderived} for the 
	helium-rich models  with $\log\mathrm{(H/He)}=-10.0$
	from \citet{cukanovaite19}. All values shown here are given in Table\,\ref{tb:Dderived-DB}.}
	\label{fg:Dderived-DB}
\end{figure}

Table\,\ref{tb:Dderived_DA_full} contains the diffusion coefficients derived analytically using Eqs.\,\eqref{eq:Dgran} and \eqref{eq:Dreal} from mean 3D properties for the grid of the warmer ($>$11\,400\,K) convective pure-hydrogen atmospheres at $\log g=8.0$ presented in \citetalias{cunningham19}. These quantities correspond to the black solid and dashed line shown in Fig.\,\ref{fg:direct_full}. In the same table and Fig.\,\ref{fg:Dderived} we show the same quantities for the grid of cooler ($<$13\,000\,K) convective DAs from \citet{tremblay13a} for $\log g=$ 7.0--9.0. The two grids have an overlap in effective temperature between 11\,400--13\,000\,K at $\log g=8.0$. In this region we include only the values for the warmer grid (i.e., that from \citetalias{cunningham19}).
We find the diffusion coefficients where the grids overlap agree to within 0.06 dex, well within the estimated uncertainty of 0.1 dex. We also present the diffusion coefficients across the grid of helium-rich models with $\log\mathrm{(H/He)}=-10.0$ published by \citet{cukanovaite19}. These are given in Table\,\ref{tab:Dderived_DB} and Fig.\,\ref{fg:Dderived-DB}.

\begin{table}
	\centering
        \caption{Diffusion coefficients derived from direct diffusion experiments using nine simulations which were adapted from \citetalias{cunningham19} and \citet{tremblay13a} for pure-hydrogen atmosphere white dwarfs at $\log g= 8.0$. 
        The horizontal diffusion coefficient is given as $D_{\rm surf}$ which represents the mean diffusion coefficient across the convectively mixed layers and is shown explicitly in Eq.\,\eqref{eq:Dhor-mean}. The associated error is the standard error on the mean. The upper limits on the diffusion coefficient (corresponding to the photospheric layers) are also given. The simulations used to derive these results included a surface gravity of $\log g=8.0$, grey opacities and boxes with 150 grid points in all three (x, y, and z) dimensions.
    }
        \begin{tabular}{llrr}
                \hline            
                \vspace{5pt}
$T_{\rm eff}$ & $\Delta(T)$ &  log \Dhor  &      log \Dhormax \\

$\mathrm{[K]}$ & $\mathrm{[K]}$ & [cm$^{2}$\,s$^{-1}$] & [cm$^{2}$\,s$^{-1}$] \\
\hline
 6000 &  +29 &   7.64 $\pm$   0.07 &   8.28 $\pm$   0.07\\
 8000 &  +41 &   8.39 $\pm$   0.09 &   8.99 $\pm$   0.07\\
10000 &  +38 &   8.92 $\pm$   0.13 &   9.57 $\pm$   0.10\\
12000 &  -17 &   9.39 $\pm$   0.12 &  10.20 $\pm$   0.12\\
13000 &   +1 &   9.26 $\pm$   0.17 &  10.02 $\pm$   0.15\\
13500 &   -2 &   8.97 $\pm$   0.16 &   9.76 $\pm$   0.13\\
14500 &   -9 &   8.38 $\pm$   0.09 &   9.21 $\pm$   0.08\\
15000 &   -1 &   8.18 $\pm$   0.03 &   8.97 $\pm$   0.06\\
16000 &   -4 &   8.10 $\pm$   0.04 &   8.85 $\pm$   0.04\\
                \hline
        \end{tabular}
        \label{tb:D3d}\\
\end{table}
\begin{table*}
	\centering
        \caption{Diffusion coefficients defined by Eq.~\eqref{eq:Dreal} for pure-hydrogen DA white dwarf atmospheres. Computed using mean 3D quantities from a grid of cool DAs ($\leqslant13\,000$\,K) from \citet{tremblay13a}, except for those with $11\,600\,\mathrm{K}\leqslant T_{\rm eff}\leqslant 18\,000$\,K and $\log g=8.0$, which were computed using the grid of deep simulations from \citetalias{cunningham19}. \change{The simulations leading to the results in this table utilised a non-grey opacity scheme for the radiative transfer}. We also show the upper limit on the diffusion coefficient (corresponding to the photospheric layers) defined by Eq.~\eqref{eq:Dgran}. The real effective temperature of a simulation can stabilise at a slightly different \change{effective temperature} to the desired value. To represent this we include the quantity $\Delta(T)$ such that the true effective temperature is given by $T_{\rm eff} + \Delta(T)$. An explicit error is not available for these results, but we note that typical \change{statistical} errors from the direct diffusion experiments (Table~\ref{tb:D3d}) were on the order of $\sigma(\log \mDhor)\approx0.1$. \change{An additional source of uncertainty arises in Fig.\,\ref{fg:direct_full} from the imperfect fit of the directly derived diffusion coefficients to those computed from physical parameters. We find at the warmest end of the grid the difference is as large as 0.2 and 0.4 dex for \Dhor\ and \Dhormax, respectively.}
    }
        \begin{tabular}{lclcccclclcc}
                 \cline{1-5} \cline{8-12}
                 \vspace{5pt}
$T_{\rm eff}$  & $\log g$ & $\Delta(T)$ & $\log$ \Dhor & $\log$ \Dhormax &   &   & $T_{\rm eff}$  & $\log g$ & $\Delta(T)$ & $\log$ \Dhor & $\log$ \Dhormax \Tstrut \\
$\mathrm{[K]}$  & [cm\,s$^{-2}$] & [K] & [cm$^{2}$\,s$^{-1}$] & [cm$^{2}$\,s$^{-1}$] &   &   & $\mathrm{[K]}$  & [cm\,s$^{-2}$] & [K] & [cm$^{2}$\,s$^{-1}$] & [cm$^{2}$\,s$^{-1}$] \Bstrut\\
				
                \cline{1-5} \cline{8-12}
               
                 7000  &   7.00 &  +46 &   9.38 &   9.88 &   &   &    13000  &   8.00 &   +4 &   9.31 &   9.97 \Tstrut\\
                8000  &   7.00 &  +27 &   9.63 &  10.19 &   &   &    13500  &   8.00 &   +1 &   9.25 &   9.84 \\ 
                9000  &   7.00 &  +25 &   9.88 &  10.47 &   &   &    14000  &   8.00 &   +1 &   9.09 &   9.65 \\ 
                9500  &   7.00 &  +21 &  10.03 &  10.64 &   &   &    14500  &   8.00 &   -2 &   8.89 &   9.41 \\ 
                10000  &   7.00 &  +18 &  10.25 &  10.88 &   &   &    15000  &   8.00 &   +0 &   8.66 &   9.16 \\ 
                10500  &   7.00 &  +40 &  10.45 &  11.13 &   &   &    15500  &   8.00 &   +1 &   8.32 &   8.81 \\ 
                11000  &   7.00 &   +0 &  10.40 &  11.09 &   &   &    16000  &   8.00 &   +2 &   7.99 &   8.50 \\ 
                11500  &   7.00 &   +1 &  10.24 &  10.80 &   &   &    16500  &   8.00 &   +3 &   7.06 &   7.54 \\ 
                12000  &   7.00 &   +1 &  10.14 &  10.64 &   &   &    17000  &   8.00 &   +3 &   6.55 &   7.44 \\ 
                12500  &   7.00 &   +1 &   9.93 &  10.40 &   &   &    17500  &   8.00 &  +24 &   5.78 &   7.02 \\ \vspace{5pt}  
                13000  &   7.00 &   +3 &   9.79 &  10.30 &   &   &    18000  &   8.00 &  +21 &   5.35 &   6.61 \\ 
                6000  &   7.50 &  +65 &   8.42 &   8.83 &   &   &      6000  &   8.50 &  +24 &   7.13 &   7.61 \\ 
                7000  &   7.50 &  +33 &   8.78 &   9.22 &   &   &    7000  &   8.50 &  -75 &   7.68 &   8.13 \\ 
                8000  &   7.50 &  +17 &   9.02 &   9.55 &   &   &    8000  &   8.50 &   +4 &   7.79 &   8.28 \\ 
                9000  &   7.50 &  +15 &   9.24 &   9.84 &   &   &    9000  &   8.50 &  +68 &   8.02 &   8.58 \\ 
                9500  &   7.50 &  +49 &   9.37 &   9.98 &   &   &    9500  &   8.50 &  +22 &   8.11 &   8.68 \\ 
                10000  &   7.50 &   +7 &   9.53 &  10.15 &   &   &    10000  &   8.50 &  -28 &   8.20 &   8.80 \\ 
                10500  &   7.50 &   +0 &   9.71 &  10.37 &   &   &    10500  &   8.50 &   -4 &   8.32 &   8.95 \\ 
                11000  &   7.50 &  -62 &   9.87 &  10.39 &   &   &    11000  &   8.50 &   -3 &   8.44 &   9.09 \\ 
                11500  &   7.50 &   -2 &   9.95 &  10.70 &   &   &    11500  &   8.50 &  -10 &   8.58 &   9.24 \\ 
                12000  &   7.50 &   -1 &   9.87 &  10.55 &   &   &    12000  &   8.50 &  -21 &   8.72 &   9.41 \\ 
                12500  &   7.50 &   +0 &   9.76 &  10.32 &   &   &    12500  &   8.50 &  -80 &   8.88 &   9.60 \\ \vspace{5pt}
                13000  &   7.50 &   +2 &   9.67 &  10.18 &   &   &   13000  &   8.50 &  -91 &   8.97 &   9.74 \\
			     6000  &   8.00 &   -3 &   7.82 &   8.28 &   &   &    6000  &   9.00 &  +28 &   6.74 &   7.28 \\ 
			     7000  &   8.00 &  +11 &   8.16 &   8.60 &   &   &    7000  &   9.00 &  -40 &   6.99 &   7.45 \\ 
			     8000  &   8.00 &  +34 &   8.41 &   8.91 &   &   &    8000  &   9.00 &  +41 &   7.19 &   7.64 \\ 
			     9000  &   8.00 &  +36 &   8.63 &   9.22 &   &   &    9000  &   9.00 &   -1 &   7.41 &   7.93 \\ 
			     9500  &   8.00 &  +18 &   8.73 &   9.32 &   &   &    9500  &   9.00 &   +7 &   7.50 &   8.05 \\ 
			    10000  &   8.00 &  +25 &   8.85 &   9.46 &   &   &    10000  &   9.00 &  -38 &   7.58 &   8.16 \\ 
			    10500  &   8.00 &  +32 &   9.01 &   9.65 &   &   &    10500  &   9.00 &  -97 &   7.66 &   8.27 \\ 
			    11000  &   8.00 &   +5 &   9.15 &   9.81 &   &   &    11000  &   9.00 &  -52 &   7.78 &   8.41 \\ 
			    11500  &   8.00 &  +29 &   9.33 &  10.02 &   &   &    11500  &   9.00 &  -85 &   7.88 &   8.53 \\ 
			    11600  &   8.00 &  +41 &   9.37 &  10.09 &   &   &    12000  &   9.00 &  -85 &   8.01 &   8.68 \\ 
			    12000  &   8.00 &   +8 &   9.45 &  10.16 &   &   &    12500  &   9.00 &  -64 &   8.14 &   8.84 \\ 
			    12500  &   8.00 &   +9 &   9.42 &  10.16 &   &   &    13000  &   9.00 &  -31 &   8.28 &   9.02 \Bstrut\\

               \cline{1-5} \cline{8-12}
        \end{tabular}
        \label{tb:Dderived_DA_full}\\
\end{table*}

\begin{table}
	\centering
        \caption{Similar to Table~\ref{tb:Dderived_DA_full} for the grid of DB white dwarfs (log [H/He] = $-$10) from \citet{cukanovaite19}.}
        \label{tab:Dderived_DB}
        \begin{tabular}{lclcc}
                \hline            
                 \vspace{5pt}
$T_{\rm eff}$  & $\log g$ & $\Delta(T)$ & $\log$ \Dhor & $\log$ \Dhormax \\
$\mathrm{[K]}$  & [cm\,s$^{-2}$] & [K] & [cm$^{2}$\,s$^{-1}$] & [cm$^{2}$\,s$^{-1}$] \\
\hline
12000  &   7.50 &  +98 &   8.10 &   8.47 \\
14000  &   7.50 &  -31 &   8.38 &   8.83 \\ 
16000  &   7.50 &  -53 &   8.65 &   9.19 \\ 
18000  &   7.50 &  +59 &   8.96 &   9.52 \\ 
20000  &   7.50 &  -69 &   9.41 &  10.09 \\ 
22000  &   7.50 &  +44 &   9.64 &  10.68 \\ 
24000  &   7.50 & -226 &   9.74 &  10.87 \\ 
26500  &   7.50 &   -3 &   9.83 &  10.93 \\ 
28000  &   7.50 &   -3 &   9.77 &  10.85 \\ 
30000  &   7.50 &   -7 &   9.63 &  10.65 \\ \vspace{5pt}
32000  &   7.50 &   +1 &   9.50 &  10.46 \\ 
12000  &   8.00 &  +20 &   7.42 &   7.75 \\ 
14000  &   8.00 &  +83 &   7.77 &   8.20 \\ 
16000  &   8.00 & +105 &   8.05 &   8.57 \\ 
18000  &   8.00 &  +82 &   8.32 &   8.89 \\ 
20000  &   8.00 &  +90 &   8.66 &   9.27 \\ 
22000  &   8.00 &   +5 &   9.06 &   9.88 \\ 
24000  &   8.00 & +144 &   9.18 &  10.23 \\ 
26000  &   8.00 & -102 &   9.25 &  10.35 \\ 
28000  &   8.00 & +107 &   9.32 &  10.39 \\ 
30000  &   8.00 &   -3 &   9.33 &  10.40 \\ 
32000  &   8.00 &   +1 &   9.30 &  10.30 \\ \vspace{5pt}
34000  &   8.00 &   +0 &   9.17 &  10.11 \\ 
12000  &   8.50 & +139 &   6.83 &   7.20 \\ 
14000  &   8.50 &   +7 &   7.15 &   7.55 \\ 
16000  &   8.50 &  -39 &   7.42 &   7.89 \\ 
18000  &   8.50 &   +0 &   7.68 &   8.23 \\ 
20000  &   8.50 &  -45 &   7.94 &   8.53 \\ 
22000  &   8.50 &   -1 &   8.33 &   9.00 \\ 
24000  &   8.50 & +143 &   8.61 &   9.53 \\ 
26000  &   8.50 & -195 &   8.67 &   9.72 \\ 
28000  &   8.50 &  -66 &   8.76 &   9.85 \\ 
30500  &   8.50 &  +67 &   8.88 &   9.95 \\ 
32000  &   8.50 & +208 &   8.88 &   9.94 \\ \vspace{5pt}
34000  &   8.50 &  +20 &   8.90 &   9.92 \\ 
12000  &   9.00 & +124 &   6.22 &   6.56 \\ 
14000  &   9.00 & +117 &   6.56 &   6.97 \\ 
16000  &   9.00 &  +29 &   6.82 &   7.28 \\ 
18000  &   9.00 &   -2 &   7.05 &   7.58 \\ 
19500  &   9.00 &  +31 &   7.25 &   7.84 \\ 
22000  &   9.00 &  -22 &   7.59 &   8.23 \\ 
24000  &   9.00 &  +82 &   7.95 &   8.71 \\ 
26000  &   9.00 & +116 &   8.10 &   9.03 \\ 
28000  &   9.00 & +143 &   8.19 &   9.23 \\ 
30000  &   9.00 & +184 &   8.27 &   9.33 \\ 
31500  &   9.00 &  -60 &   8.32 &   9.38 \\ 
34000  &   9.00 & +105 &   8.43 &   9.46 \\
                \hline
        \end{tabular}
        \label{tb:Dderived-DB}\\
\end{table}

\subsection{Surface spreading}
\label{sec:results-surface-spreading}

{In the following we consider two idealised geometries of accretion. Firstly we examine \textit{instantaneous accretion} onto a point source. In nature this could arise from, for instance, the direct impact of an asteroid. Probabilistic arguments suggest that this scenario could provide on the order of 1--2 per cent of metals in white dwarfs. Nonetheless it allows for a highly idealised problem setup and bears relevance for the decreasing phase at the end of any steady-state accretion.} 

{Secondly, we consider the more likely scenario of \textit{continuous accretion} onto a point. This would correspond to the canonical model of debris from a disc constantly feeding the white dwarf surface. Both models can be scaled to account for equatorial accretion, rather than accretion onto a spot. If the white dwarf possesses a sufficiently strong magnetic field, material may be channelled toward the poles \citep{metzger12}. This scenario is well described by our second model, where material is constantly fed to two spots, rather than one. We revisit the importance of magnetic fields regarding the delivery of material to the surface in Section \,\ref{sec:mag} and the impact of magnetic fields on convective instabilities in Section\,\ref{sec:accretion-geometry}. In the following we focus our modelling on the transport of material due to convective motions after it arrives at the surface.}
	
\subsubsection{Instantaneous accretion}
We begin by considering the evolution of a system whereby a large quantity of a passive scalar is instantaneously delivered either to a point on the white dwarf surface or equatorially, e.g. from an accretion disc.

In the first model, to account for the large range of timescales involved, we consider the spreading and sinking to be governed by two independent, one-dimensional equations. One governs the horizontal spreading of material and assumes the {passive scalar concentration is} restricted to the convection zone. {The other equation relaxes the massless assumption, allowing the accreted material to sink via microscopic diffusion processes; i.e.,  gravitational settling.} 
Clearly this problem setup is a strong simplification but it allows for a simple analytic formulation of the problem. The analytic approach refers to the problem setup outlined in Section\,\ref{sec:1Dmodel}.

\begin{figure}
 \centering
 \subfloat{\includegraphics[width=1.\columnwidth]{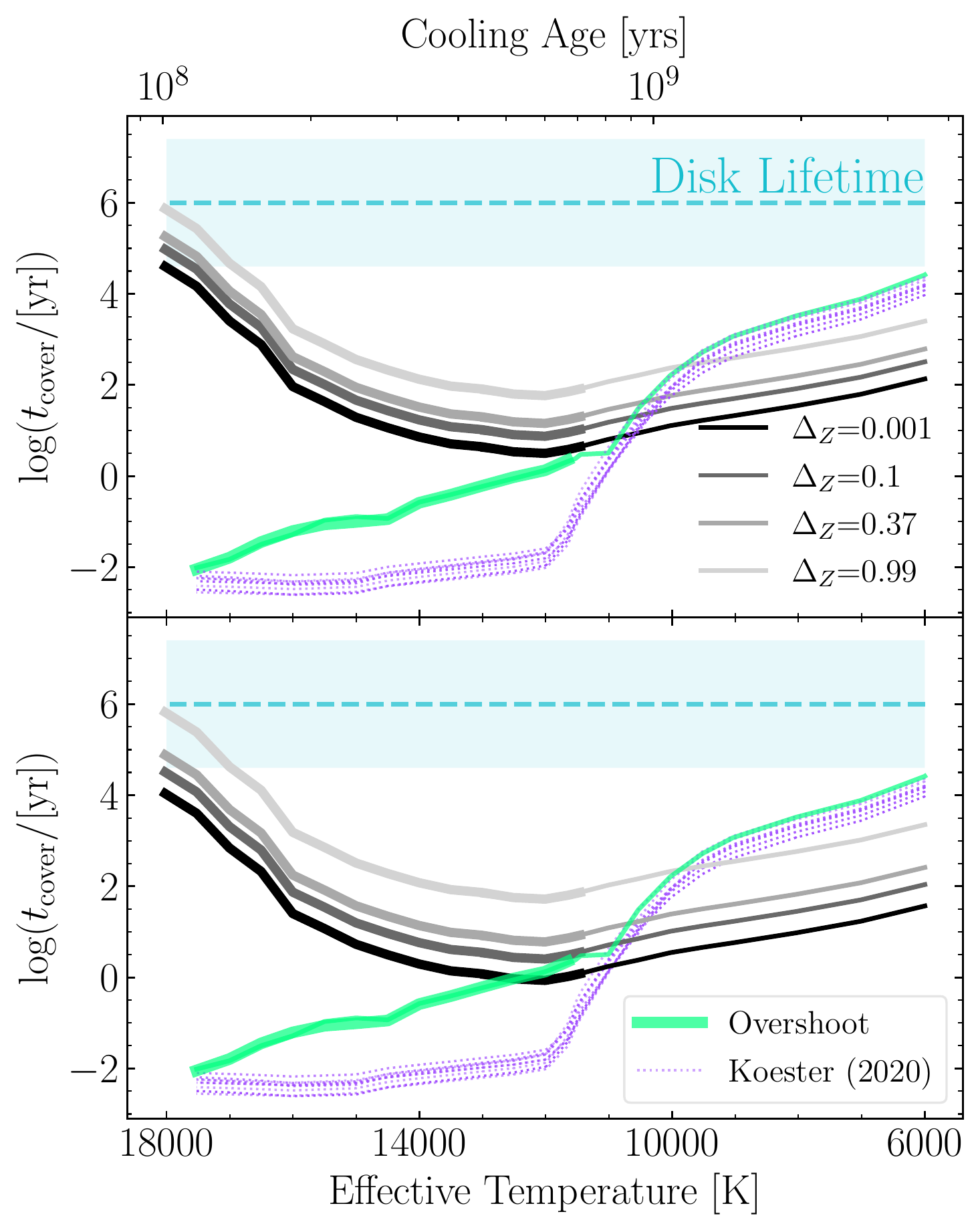}}
 \caption{Estimates of the timescales, for DA models with $\log g=8.0$, corresponding to key physical processes in our model. From black to grey are the times required for the white dwarf surface to homogenize such that the variation of metal abundance with position, $\Delta_Z$, is less than 0.001, 0.1, 0.5 and 0.99, respectively. The top panel depicts $t_{\rm cover}$ for instantaneous accretion onto a point whilst the bottom panel assumes instantaneous equatorial accretion. The diffusion calculations of \citet{koester2020} yield the vertical sinking times for all elements with $Z<31$. The sinking time scales for C, O, Na, Mg, Al, Si, Ca and Fe are shown in purple dotted for clarity. The modified sinking times inclusive of convective overshoot are shown in green. The overshoot correction in the range 18\,000--11\,400\,K is taken from \citepalias{cunningham19} (thick) and is computed as the mean correction over elements Fe, Ca and Mg. The overshoot correction for cooler models is estimated to be an additional one pressure scale height of mixing below the lower Schwarszchild boundary (thin). The dashed cyan line depicts an estimate of a disc lifetime and the expected uncertainty (see Table\,\ref{tab:lifetimes}). {This figure highlights the strong, effective temperature dependence of the characteristic time scales of horizontal spreading and vertical sinking. The minimum in $t_{\rm cover}$ around 12\,000\,K corresponds to the maximum in the horizontal convective velocity at that temperature (see Fig.\,\ref{fg:grid_overlap}, top panel).} The cooling ages shown on the upper x-axis of this and subsequent figures correspond to those from the cooling models of \citet{fontaine01} for white dwarfs with mass $M=0.6\,M_{\odot}$ and a thick H-envelope.}
 \label{fg:timescales}
\end{figure}

Fig.\,\ref{fg:timescales} shows the time required, $t_{\rm cover}$, for the metal abundance at the white dwarf surface to homogenize. From black to grey this is shown for increasing levels of homogeneity; namely contrasts of $\Delta_Z=0.001,0.1,0.37,0.99$. The vertical diffusion (sinking) times from \citet{koester2020} are shown in dotted purple for eight elements across the full effective temperature range. The mean sinking time arising from the model of convective overshoot \citepalias{cunningham19} is shown in green for effective temperatures between 11\,400--18\,000\,K. For comparison our updated estimate of a typical disc lifetime 
is shown in dashed blue, with the associated error (see Section\,\ref{sec:disc-lifetimes}). We find moderate dependence on whether the accretion is spot-like or equatorial from this model. The expected behaviour that equatorial accretion homogenizes faster is recovered, but the difference is no more than a factor of three. 

\begin{figure}
 \centering
 \subfloat{\includegraphics[width=1.\columnwidth]{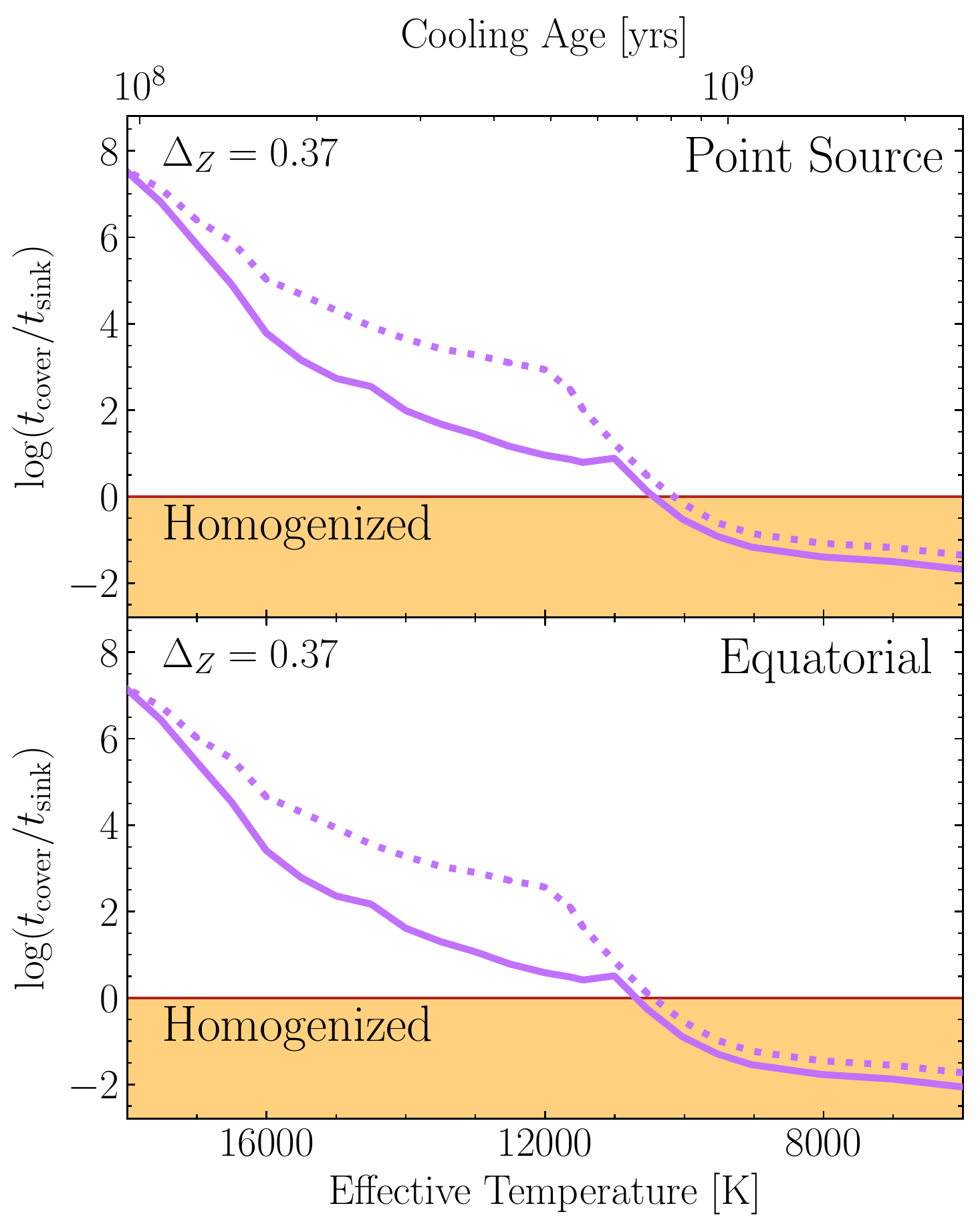}}
 \caption{The ratio of the spreading and sinking timescales, $t_{\rm cover}$ and $t_{\rm sink}$, is shown here in purple across the range of effective temperatures considered in this study for H-rich atmospheres with $\log g=8.0$. As with Fig.\,\ref{fg:timescales}, the top and bottom panels correspond to accretion onto a spot and accretion onto the equator, respectively. The red horizontal line indicates where the timescales are equal. 
 Dotted lines correspond to the 1D vertical diffusion times of \citet{koester2020} while thick solid lines are the results of \citetalias{cunningham19} taking into account convective overshoot. The spreading timescale is defined for a maximum abundance contrast of $\Delta_Z=0.37\approx1/e$.}
 \label{fg:homogenize01}
\end{figure}

We introduce in Fig.\,\ref{fg:homogenize01} the ratio of the spreading time, $t_{\rm cover}$,
and the sinking time, \tsink. This dimensionless quantity is a metric by which we will assess the capability of a white dwarf envelope to efficiently spread material. One can consider that when this ratio is below unity the surface layers are able to spread material out quicker than it sinks. In this figure we show the $t_{\rm cover}/t_{\rm sink}$ ratio for $t_{\rm cover}$ corresponding to $\Delta_Z=1/e\approx0.37$, i.e., chemical abundance variations of less than a factor $\approx$3.

\begin{figure}
 \centering
 \subfloat{\includegraphics[width=.8\columnwidth]{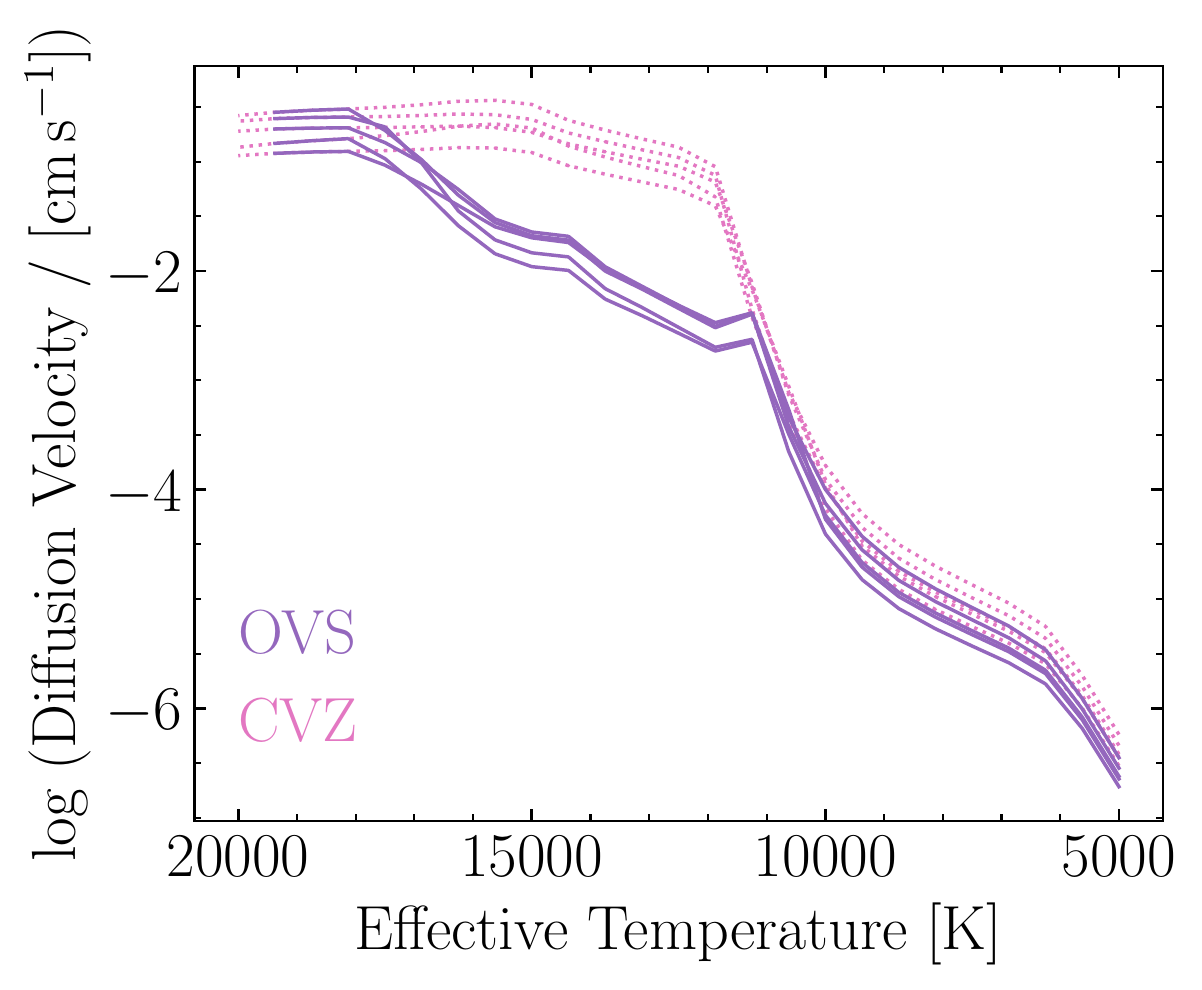}}
 \caption{Vertical diffusion velocities for O, Mg, Si, Ca and Fe defined at the base of the mixed region for H-rich atmospheres with $\log g =8.0$. 
 	Diffusion velocities at the lower Schwarzschild boundary (CVZ) taken from updated tables of \citet{koester2020} and are shown in pink (dotted). The diffusion velocities presented in \citetalias{cunningham19} which account for significant convective overshoot regions (OVS) between 14\,000 and 18\,000\,K are shown in purple (solid). For lower effective temperatures ($<11\,400$\,K) the diffusion velocity is taken one pressure scale height beneath the lower Schwarzschild boundary as a minimum estimate of the contribution of convective overshoot. For higher effective temperatures with radiative atmospheres ($>18\,000$\,K)  the diffusion velocity is instead given at an optical depth of $\tau_{\rm R}\sim 1$.}
 \label{fg:vdiff_koester09}
\end{figure}

\begin{figure}
 \centering
  \subfloat{\includegraphics[width=1.\columnwidth]{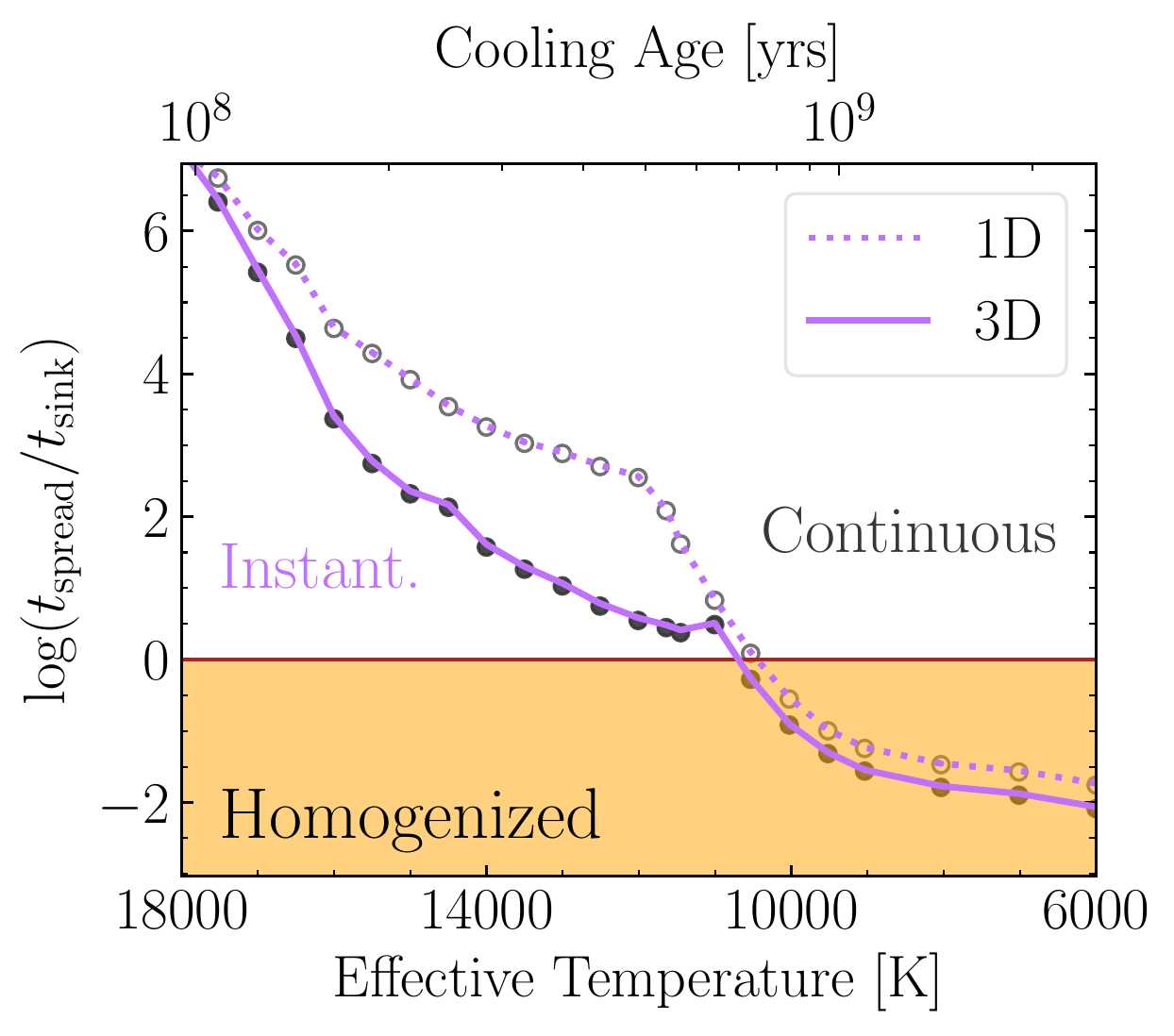}}
 \caption{Comparison of two models, instantaneous accretion (purple lines) and continuous accretion (grey circles), in the comparison of spreading and sinking timescales for H-rich atmospheres at $\log g=8.0$. The instantaneous accretion model is shown where the vertical sinking velocity, $v_{\rm drift}$, is either given by the 1D mixing length theory results of \citet{koester2020} or from the 3D overshoot results of \citetalias{cunningham19} (see Fig.\,\ref{fg:vdiff_koester09}) in dotted or solid, respectively. The spreading timescale is defined for a maximum surface abundance contrast of $\Delta_Z=0.37$ from equatorial accretion (reproduced from lower panel; Fig.\,\ref{fg:homogenize01}). The latter model -- continuous accretion -- is expressed by the left-hand side of Eq.\,\eqref{eq:tsinkspreadConstant} as a mean across the elements O, Mg, Si, Ca and Fe. The overshoot and no overshoot description of the continuous accretion case is indicated by filled and open circles, respectively.
 }
 \label{fg:homogenise05-inst.-cont.}
\end{figure}

\subsubsection{Continuous Accretion}
\label{sec:continuous-accretion}
So far we have considered the scenario that the accreted metals arrive in a small region on the surface at one instant in time. This would correspond to a direct impact type scenario and is considered one of the least likely sources of accretion at white dwarfs \citep{wyatt14}. We have also required that the metals remain confined to the surface when modelling the spreading. We now relax these criteria and go on to consider the scenario in which a white dwarf accretes for a prolonged period with constant accretion rate. This would represent the stable accretion of material from a debris disc.

To model the behaviour of the metals at the surface of the white dwarf in such a system we adopt a two-zone model. In this model the horizontal diffusion coefficient, $\mDhor$, is confined to some upper region and a lower adjacent region has a drift velocity in the vertical direction, $v_{\rm drift}$. {This describes the physical scenario of a trace concentration being removed from a mixed region at the surface with a drift velocity in the vertical direction defined at the base of said region.} 

The equation describing the evolution of this scenario is as follows
\begin{equation}
    \frac{\partial u}{\partial t} = (\mDhor) \nabla^2 u - v_{\rm drift}\frac{\partial u}{\partial z} \, ,
    \label{eq:2d-two-zone}
\end{equation}
where $u=u(x,y,z,t)$ is the concentration of a passive scalar depending on depth, surface position and time. Once this system has reached a steady state, the equation reduces to
\begin{equation}
    (\mDhor) \nabla^2 u = v_{\rm drift} \frac{\partial u}{\partial z}\, .
\end{equation}
For a constant flux delivered to $x = y = 0$, this equation has the analytic solution of the form
\begin{equation}
    u(x,y,z) \propto z^{-1/2} \exp \left( \frac{-(x+y)^2}{Az} \frac{v_{\rm drift}}{\mDhor} \right) \, ,
    \label{eq:diff-eqn-continuous}
\end{equation}
where $A=8$ in 3D and $A=4$ in 2D ($y=0$), where $z$ is the vertical extent of the surface zone, $v_{\rm drift}$ is the vertical diffusion velocity at the base of the surface zone and $x$ is the horizontal extent of the surface. We are interested in the global spreading properties which for a spot would require $x=y=\pi R_{\rm WD}$ and for an equatorial belt would require $x=y=\pi R_{\rm WD}/2$. Thus in the case of a white dwarf with equatorial accretion, for convection to be efficient at spreading before material sinks requires
\begin{align}
\frac{(\pi R_{\rm WD}/2)^2}{4z_{\rm cvz}}\frac{v_{\rm drift}}{\mDhor} &\lesssim 1 \, ,
    \label{eq:tsinkspreadConstant}
\end{align}
where $z_{\rm cvz}$ is the depth of the convection zone, $v_{\rm drift}$ represents a characteristic velocity for the settling of a trace amount of metal out of the convection zone. This can also be written
\begin{equation}
	\frac{t_{\rm spread}}{t_{\rm sink}} \lesssim 1 \, ,
	\label{eq:tspread/tsink}
\end{equation}
where the transport timescales are estimated as
\begin{align}
	t_{\rm spread} &\equiv \frac{(\pi R_{\rm WD}/2)^2}{4\mDhor} \, ,	\label{eq:tspread} \\
	t_{\rm sink}     &\equiv \frac{z_{\rm cvz}}{v_{\rm drift}} \, .	\label{eq:tsink}
\end{align}

\noindent For the drift velocity we adopt the {vertical} diffusion velocity at the base of the mixed region. Fig.\,\ref{fg:vdiff_koester09} shows the diffusion velocity for O, Mg, Si, Ca and Fe computed at the lower Schwarszchild boundary (pink-dotted) and at the base of the overshoot region (purple-solid) taken from the {diffusion calculations of \citet{koester2020} and overshoot determination of \citetalias{cunningham19}}, respectively. As was shown by the latter, convective instabilities arise at 18\,000\,K.  In the radiative regime ($T_{\rm eff} > 18\,000$\,K) diffusion velocities are given at an optical depth of $\tau_{\rm R}\sim 1$. We caution that none of these models include the effects of radiative levitation and thermohaline mixing which may become important in the radiative regime \citep{chayer95a,chayer95b,bauer18}.

We plot the left-hand side of Eq.\,\eqref{eq:tsinkspreadConstant} in Fig.\,\ref{fg:homogenise05-inst.-cont.} in grey circles and we include the results with (solid) and without (open) overshoot. We also reproduce the results for instantaneous accretion (see Fig.\,\ref{fg:homogenize01}) in purple, again with (solid) and without (dotted) overshoot. {We find the two models make almost entirely overlapping predictions on the ability of a white dwarf to homogenize its surface within a diffusion timescale.}

The models we have presented highlight the horizontal mixing capabilities across the full range of convective H-atmosphere white dwarfs. In Figs.\,\ref{fg:direct_full}~\&~\ref{fg:Dderived} and Table\,\ref{tb:Dderived_DA_full} we provide the horizontal diffusion coefficients for all convective pure-H atmosphere white dwarfs with surface gravities $\log g=$ 7.0--9.0. For brevity, in this study we have only compared those at $\log g=8.0$ with vertical diffusion coefficients (Figs.\,\ref{fg:timescales}, \ref{fg:homogenize01}~\&~\ref{fg:homogenise05-inst.-cont.}), as the results are similar for other surface gravities but with a small shift in effective temperature. The effect of trace metals in H-rich model atmosphere calculations is very small or negligible, hence our results directly apply to DAZ white dwarfs. We now go on to consider the surface spreading behaviour in helium-atmosphere white dwarfs.

\subsubsection{Surface spreading in He-atmosphere white dwarfs}
\begin{figure}
 \centering
 \subfloat{\includegraphics[width=1.\columnwidth]{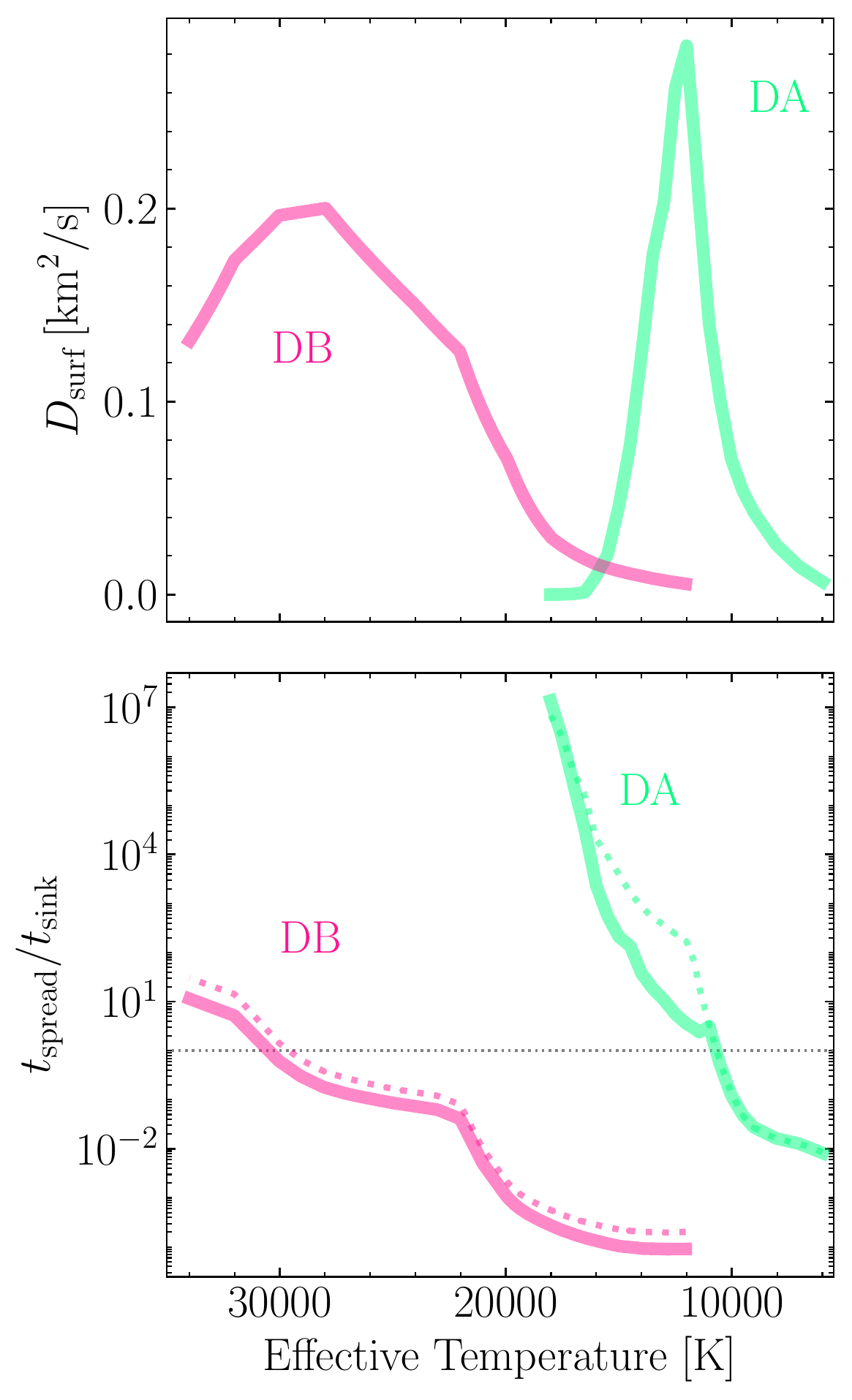}}
 \caption{{\it Top:} Horizontal diffusion coefficients for DA (green) and DB (pink) models derived from mean 3D parameters from Eq.\,\eqref{eq:Dreal} and the grids published in \citet{cunningham19,tremblay13a} and \citet{cukanovaite19}, respectively. {\it Bottom:} The ratio of the spreading time and sinking times (Eq. \ref{eq:tspread/tsink}) for DA (green) and DB white dwarfs (pink). The dotted lines show the ratio for sinking times  defined at the lower Schwarzschild boundary. The solid lines include the correction for convective overshoot.
}
 \label{fg:ratio-DA-DB}
\end{figure}

Approximately half of the known polluted white dwarfs {have helium-rich atmospheres} \citep{farihi16}. In this section we repeat some of the procedures detailed in the previous sections to provide a picture of the mixing capabilities of He-atmosphere white dwarfs as they cool. The lower panel of Fig.\,\ref{fg:ratio-DA-DB} shows the ratio of the spreading and sinking times for our 3D grids of pure-He (pink) and pure-H (green) model atmospheres. A 3D correction for convective overshoot is not readily accessible for the DBs -- requiring an overshoot study similar to \citetalias{cunningham19} for the early convective DBs.
This is outside the scope of this work, where we instead adopt an arbitrary overshoot correction of one pressure scale height of additional mixing below the lower Schwarzschild boundary.
A comparison of the solid and dotted pink lines in Fig.\,\ref{fg:ratio-DA-DB} shows that this assumption does not significantly impact the results of the pure-He models. Given that the majority of metal-polluted DBs have effective temperatures less than 25\,000 K, this figure suggests that metal-polluted DBs are far more likely to be homogeneously mixed within a diffusion timescale. While there are no 3D simulations of helium-rich atmospheres cooler than 12\,000\,K (DC, DZ and DQ white dwarfs), diffusion timescales calculated from 1D model atmospheres suggest that $t_{\rm spread}/t_{\rm sink}$ plateaus at values in the range 10$^{-2}$--10$^{-3}$, depending on trace hydrogen and metal abundances.

\section{Application to white dwarf population}
\label{sec:discussion-WDpop}
For any white dwarf, the numerical solution of Eq.\,\eqref{eq:2d-two-zone}, using the horizontal diffusion coefficients presented in this paper as well as vertical diffusion coefficients from \citetalias{cunningham19} or the literature, would yield a reasonable description of the distribution of metals across the white dwarf surface. This could provide the basis for the generation of synthetic spectra under different accretion conditions (geometries and time-dependence) and rotational periods. In Section\,\ref{sec:geometric-syn-spectra} we generate synthetic spectra for the DAZ white dwarf SDSS J104341.53+085558.2, assuming the observed metals are confined to varying spot sizes on the visible surface. In the following, 
we firstly explore a heuristic discussion of these systems across a wide parameter space. 

The results we have shown so far focus mostly on the competition between {horizontal} spreading and {vertical} sinking. This led to a separation, based on atmospheric parameters, between metal-rich white dwarfs which are expected to have a homogeneous surface composition and those {where metals must be concentrated in and near regions of accretion.} 
When it comes to predict how likely it is to observe photometric or spectral variability in a white dwarf, we also need to consider the amplitude, periodicity and total duration of a predicted contrast, which depend on further physical variables. In addition to (1) the white dwarf atmospheric parameters (spreading and sinking times), we must also consider (2) the geometry of accretion, (3) the instantaneous accretion rate, (4) the duration of the event (e.g. the disc lifetime combined with the sinking time), and (5) the white dwarf rotation period and inclination with respect to the line of sight. Fortunately, some of these quantities are observationally or theoretically constrained and limiting cases are sometimes sufficient when considering the white dwarf population as a whole.

\subsection{Variables}
\subsubsection{Accretion geometry}
\label{sec:accretion-geometry}

To predict the time evolution of surface inhomogeneities on white dwarfs,
one should consider the initial distribution of metals onto the white dwarf surface, whether that be a spot or multiple spots as seen in the Shoemaker-Levy 9 disruption and accretion of an asteroid onto Jupiter \citep{brown2017}, a stable accretion disc funneling material either to an equatorial belt or to the poles \citep{wyatt14,farihi2018XRay}, or a near-homogeneous (spherical) deposition. We emphasise that we obtain a trivial result if accretion is fully spherical, in which case the surface is homogeneous and diffusion is only of relevance in the radial direction, where 3D convective effects nevertheless still play a role. We will consider the relevance and motivation for each of these scenarios in the following.

\vspace{5pt}
{\noindent \textit{Direct impacts}: 
	To understand how prevalent direct impacts are as a source of observed metal pollution one must consider the likelihood of an asteroid colliding directly with the white dwarf surface. A key input to this problem is whether an asteroid can survive the tidal forces as it approaches the white dwarf. This can be quantified with the Roche radius which describes the distance at which a substellar body, held together by its own self-gravity, will succumb to the tidal forces of the white dwarf and tidally disrupt. From \citet{veras2017}, the Roche radius for a white dwarf can be written (Equation 1 of \citealt{bear2013})
\begin{equation}
	R = 0.65R_{\odot}C\left( \frac{M_{\rm WD}}{0.6M_{\odot}}\right)^{1/3} \left(\frac{\rho}{3\,\mathrm{g\,cm^{-3}}}\right)^{-1/3} \, ,
	\label{eq:roche}
\end{equation}
where $\rho$ is the density of the substellar body and $C=1.3$--2.9 accounts for additional variables not explicitly included here such as strength and shape.} 

If a given substellar body is assumed to be a strengthless rubble pile, the Roche radius is typically $R_{\rm Roche}\sim100\,R_{\rm WD}$. This implies that direct impacts are unlikely to contribute more than 1--2 per cent of metal pollution at white dwarfs \citep{wyatt14}.
If the substellar body has internal strength, however, the Roche radius is decreased \citep{veras16} and such high-strength objects can survive many passes within the rubble pile Roche radius of the white dwarf. For instance, the planetesimal observed at SDSS J1228+1040 \citep{manser2019} 
lies at 0.73 $R_{\odot}$, which is 25 per cent within what is typically considered the white dwarf's rubble pile Roche radius.
The existence of such a system provides evidence that solid bodies can exist within the Roche radius, and thus direct impacts cannot be ruled out as unphysical. 
However, the frequency of metal-pollution in white dwarfs with short sinking times (i.e., hydrogen-rich DAZ white dwarfs) strongly supports a model of (continuous) accretion from a long-lasting source, such as an accretion disc.

\vspace{5pt}
\noindent\textit{Accretion discs:} If a substellar body comes within a Roche radius and does not result in a direct impact it most likely suffered tidal disruption, forming a disc of dust and gas. There is significant evidence that the bulk of metal pollution at white dwarfs arrives this way. For one, {warm} DAZs have sinking time scales on the order of days to years, so the likelihood of observing them is low, unless the source of material has longevity. Furthermore, the observed frequency of IR excess due to dust at metal-polluted white dwarfs confirms the importance of this scenario \citep{rocchetto15}. Once a disc is formed it is likely to accrete (quasi-)continuously driven by drag forces \citep{rafikov2011,metzger12}. This provides the ongoing source of material at white dwarfs, but constraints on where the material arrives on the stellar surface are inexistent.

It has been shown that magnetic fields are likely to play an important role in the trajectory of material between the disc and the white dwarf surface. The \alfven\ radius describes the distance at which the magnetospheric and disc stresses equalize, allowing a surface magnetic field to dynamically alter the trajectory of in-falling material. If the surface magnetic field is dipolar, the \alfven\ radius for a gas disc accreting onto a white dwarf 
can be written \citep{ghosh1978} as
\begin{align}
R_{\rm A} &\approx \left( \frac{3B_{\rm WD}^2R_{\rm WD}^6}{2\dot{M}\sqrt{GM_{\rm WD}}}  \right)^{2/7} \, ,
\label{eq:ra1}
\end{align}
{where $B_{\rm WD}$ is the surface field strength and $\dot{M}$ the accretion rate.}
For a white dwarf with mass 
$M_{\rm WD}=0.6\,M_{\odot}$ and radius $R_{\rm WD}=9.0\times 10^8$\,cm this becomes
\begin{align}
&\approx 0.52 R_{\odot} \left( \frac{B_{\rm WD}}{\mathrm{kG}}\right)^{4/7} \left( \frac{\dot{M}}{10^{10}\mathrm{g\,s^{-1}}}\right)^{-2/7} \, .
\label{eq:ra2}
\end{align}

\noindent If the \alfven\ radius is smaller than the white dwarf radius ($R_{\rm A} < R_{\rm WD}$) then accretion from a disc is most likely to land in an equatorial belt on the white dwarf surface. On the other hand, if $R_{\rm A} > R_{\rm WD}$, then any material within the \alfven\ radius, and external to the white dwarf, is likely to be magnetically driven, i.e., towards the magnetic poles.
\citet{metzger12} showed that fields as weak as 0.1--1\,kG were sufficient to affect the flow of gas near the sublimation radius around white dwarfs. The authors pointed out two caveats to this prediction. Firstly, current observations constrain the total surface magnetic field, of which the dipole may only be a minor component. And secondly, debris in actively accreting systems could be diamagnetically screened, essentially exposed to systematically lower magnetic fields.

Another geometric possibility is accretion through multiple misaligned discs/rings, only a fraction of which are within our line-of-sight. Asteroids scattered to the Roche radius are highly unlikely to all be in the same plane. In fact, they could easily vary in inclination. \citet{mustill2018} showed that the orbital inclinations of bodies as they cross the Roche limit are roughly isotropic. Whilst we have only modest constraints on the accretion geometry at metal-polluted white dwarfs, a comparison can be drawn with a better studied class of accreting white dwarf system; cataclysmic variables (CVs).

\vspace{5pt}
\noindent {\it Analogy to accretion on CVs:} The hypothesis of magnetically-driven accretion is observationally confirmed for the highly-magnetic CVs (polars) such as AM Herculis \citep{gaensicke2001,schwope2020}. 
But although they provide the closest, well-studied analogy to debris-accreting white dwarfs, {it is important to understand that the range of accretion rates of $10^{7}-10^{11}$\,g\,s$^{-1}$ estimated for metal-polluted white dwarfs (Fig.\,\ref{fg:JF-disc-acc}) is 5--10 orders of magnitude lower than that observed in CVs.}

In the equatorial accretion belt scenario, material is slowed down from Keplerian velocities in the inner disc to the rotation velocity of the white dwarf. Since the free-fall velocity for a white dwarf is $\sim$1000\,km\,s$^{-1}$, compared to rotational velocities of the order of 1--10\,km\,s$^{-1}$, that must involve significant shear, turbulence, shocks and mixing above the white dwarf atmosphere. For CVs, \citet{patterson85} found that if the accretion rate is sufficiently high ($\dot{M}>10^{16}$\,g\,s$^{-1}$) the accretion disc may survive all the way down to the white dwarf surface. 
For lower accretion rates ($\dot{M}<10^{16}$\,g\,s$^{-1}$), still orders of magnitude above the range observed for metal-polluted white dwarfs, they found that the disc was likely to evaporate before reaching the white dwarf surface, forming a more diffusive equatorial belt, {or even a spherical shroud}, covering much larger latitudes. A theoretical study by \citet{meyer1994} considered the accretion disc around dwarf novae. They found that a siphon flow is capable of evaporating some of the inner disc, leading to a hole between the white dwarf surface and the inner disc. These studies support the idea of close-to-spherical accretion for CVs with low accretion rates ($\dot{M} < 10^{16}$\,g\,s$^{-1}$). However, \citet{piro} found the mixing region to be relatively small and restricted to less than one degree of the equator.

\subsubsection{Modelling disc lifetimes}
\label{sec:disc-lifetimes}
With enhanced mixing from convective overshoot \citepalias{cunningham19}, our new diffusion models
can be used to revisit and update a known issue with metal accretion rates and duty cycles across the population of 
all polluted white dwarfs, especially the distinct behaviour of H- versus He-rich atmospheres.  It has been known for some time that, all else being
equal, if identical calculations are made for instantaneous accretion rates onto known DAZ stars and known DBZ or similar He-rich stars, there 
is a sizable gap between the two inferred rates for each sub-population \citep{girven12}.  Based on the new results presented in this work and in \citetalias{cunningham19}, these 
accretion rate inferences, and their implications are updated.

H-atmosphere white dwarfs permit instantaneous accretion rate inferences based on their relatively short diffusion timescales, which reduce the full
evolutionary accretion behaviour to a simple equation (\citealt{jura09}, Equation 1).  This is not the case for He-dominated atmospheres 
because their sinking timescales are sufficiently long that the full differential equation is not immediately solvable without a knowledge of the full history.  
Nevertheless the same, simple equation can be applied to better understand the population as a whole, and to put both H- and He-atmosphere stars onto 
the same footing \citep{farihi09}.  Previous work has shown that when this is done carefully with the most up-to-date models at the time, and 
assuming the same fractional representation of metal abundance by primary atmospheric constituent, there is a notable difference where some of the He
atmosphere stars exhibit accretion rates that are orders of magnitude larger than the largest H-atmosphere accretion rates \citep{girven12}.

Assuming that the accretion rate is constant and that the accretion and diffusion (atmospheric settling) processes have reached a steady state, the rate of accretion can be estimated following \citet{koester09} as
\begin{equation}
\dot M_z = \sum\limits_{i} \frac{X_{i}M_{\rm cvz}}{\tau_i}~,
\end{equation}
{where $X_i$ is the atmospheric abundance of element $i$, $M_{\rm cvz}$ is the mass of the convectively mixed surface layers and $\tau_i$ is the diffusion timescale at the base of the mixed region.}
It must be kept in mind that these are not true accretion rates, but an average over the past sinking timescale, where the calculation simply takes the
mass of metals currently in the mixing layer and divides by a characteristic sinking timescale.  Nevertheless, they are informative in at least two
ways.  The first is in the numerator of this calculation, which represents the total mass of metals currently within the outer, visible and well-mixed layers
of the star.  The difference between the H- and He-atmosphere stars indicates that metals continue to accumulate far longer than a typical sinking timescale 
for H-atmosphere white dwarfs, and this difference has been used to calculate a typical disc lifetime \citep{girven12}.  Specifically, if one takes the larger
mass of metals that can be inferred within He-atmosphere stars, and divides by the instantaneous accretion rate inferences for H-atmosphere stars, the outcome
has dimensions of time, which can be interpreted as a disc lifetime, if the accretion rate is constant over a typical He-atmosphere sinking timescale.

The second difference is in the accretion rates themselves.  When the ongoing accretion rates for H-atmosphere stars are compared with the time-averaged
accretion rates for He-atmosphere stars, there is a similar, stark difference, and the only hypothesis in the literature suggests this is likely due to highly variable
accretion rates \citep{farihi12}.  The idea is straightforward:  only modest accretion rates are seen to be ongoing currently via H-rich white dwarfs,
but over a typical, Myr timescale for metal diffusion in He-rich white dwarfs, there have been high-rate bursts that are not yet witnessed as ongoing.
This result implies the empirical disc lifetime estimates made from these calculations are uncertain at best, but it does provide a constraint on the 
duty cycle of high- and modest-rate accretion for the population of white dwarfs as a whole, from which further tests might be made.

\begin{figure}
	\centering
	\subfloat{\includegraphics[width=1\columnwidth]{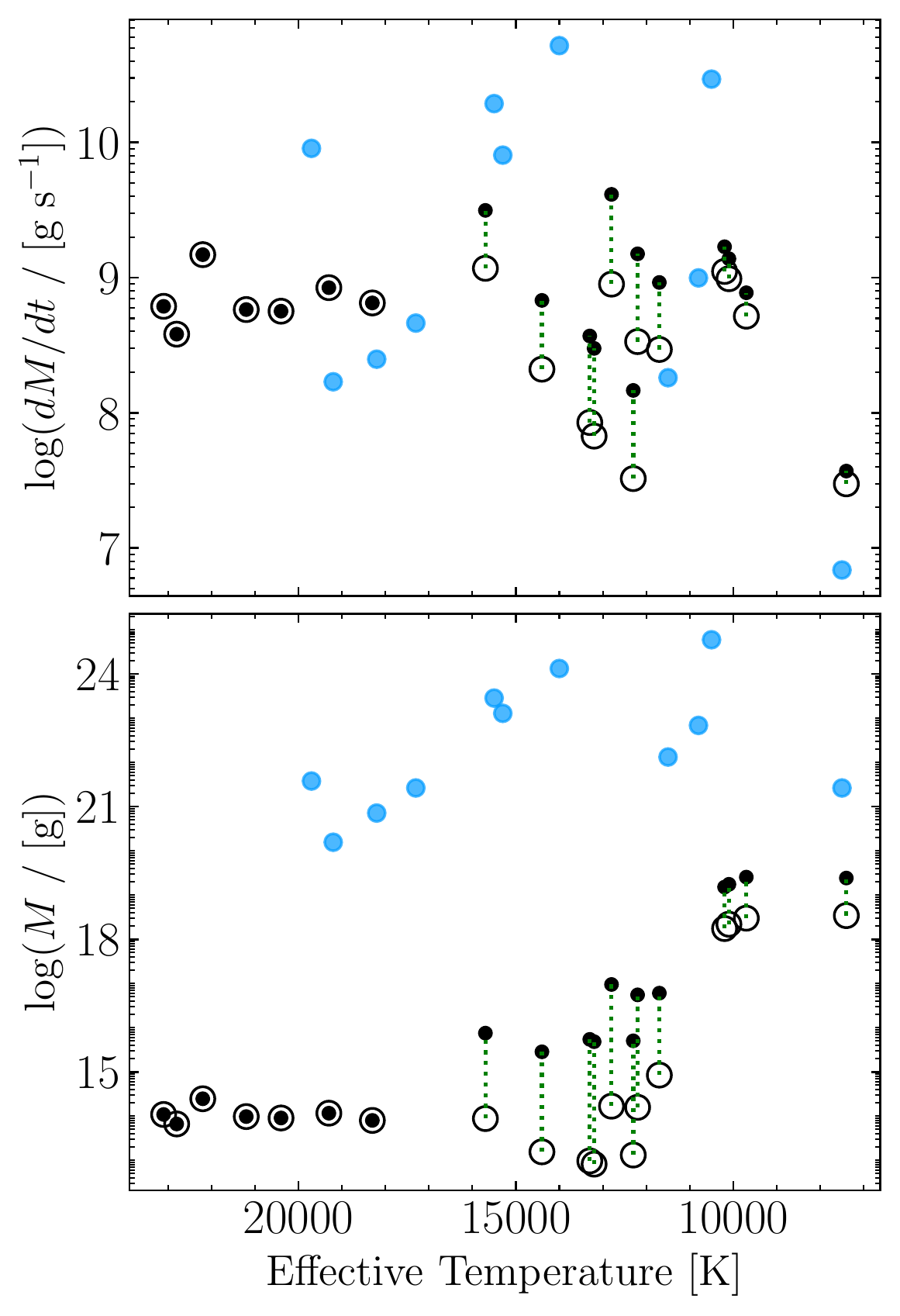}}
	\caption{Accretion rates (top panel) and lower limit on the accreted mass (bottom panel) for a sample of 19 H-rich DAZ (open, black) and 11 DBZ/DZ (solid, blue) white dwarfs with a detected IR excess from \citet{farihi16}. The overshoot correction for the DAZs is shown (solid, black), and connected to the no-overshoot case with dotted green lines for clarity.}
	\label{fg:JF-disc-acc}
\end{figure}

In Fig.\,\ref{fg:JF-disc-acc}, the new model calculations provide updated accretion rates and accreted masses for white dwarfs with confirmed IR excess. It is most noteworthy that they corroborate the previous
findings discussed above.  That is, there remains a gap between the time-averaged accretion rates for He-atmosphere stars, and the instantaneous accretion
rates for H-atmosphere stars. Table\ref{tab:lifetimes} provides the mean accretion rates for DAZ stars and accreted masses for DBZ stars with confirmed IR excess, values which we will use in the following discussion regarding disc lifetimes.

\citet{girven12} derived a crude estimate for the lifetime of a disc of disrupted planetesimals around a white dwarf to be 
\begin{equation}
t_{\rm disc} \sim \frac{\MDBZ}{\MdotDAZ}\, ,
\end{equation}
 where $\MdotDAZ$ is the mean inferred accretion rate at warm polluted hydrogen-rich DAZ stars  
 and $\MDBZ$ is
 the mean mass of metals enclosed in He-rich (DBZ and DZ) white dwarf convection zones, with both quantities derived from a sample of white dwarfs with confirmed IR excess. As discussed before, this calculation assumed a Jura-like disc model and a constant accretion rate over the full sinking timescale. In these calculations it is estimated that Ca makes up 1.6 per cent of the accreted material, a value characteristic of the bulk Earth abundance.
\citet{girven12} found for the thirteen DAZ white dwarfs with IR excess confirmed with \textit{Spitzer} the mean accretion rate to be $\MdotDAZ=9.7\times10^8\,\mathrm{g\,s^{-1}}$. The average total accreted mass for the eight DBZ/DZ stars with confirmed IR excess was found to be $\MDBZ=4.1\times 10^{22}$\,g. The accretion rates and accreted masses span orders of magnitude in range (see Fig.\,\ref{fg:JF-disc-acc}), and thus the authors favoured a logarithmic mean, such that the mean accretion rate in the DAZ stars was found to be $\lgMdotDAZ = 8.8$ [g s$^{-1}$] and mean total accreted mass in the DBZ/DZ stars $\lgMDBZ = 21.9$ [g]. This led to a disc lifetime estimate of $\log(t_{\rm disc}) = 5.6 \pm 1.1$ [yr].

\begin{table*}
	\centering
	\caption{Mean parameters for white dwarfs with inferred accretion rates and confirmed IR excess. Accretion rates and accreted masses without overshoot were taken from \citet{farihi16} for a sample of 19 H-rich (DAZ) and 11 He-rich (DBZ/DZ) metal-polluted white dwarfs with a detected IR excess in \textit{Spitzer} observations. For the the DAZ white dwarfs, these quantities were also updated with the overshoot correction from \citetalias{cunningham19}. The accretion rates and accreted masses used to derive the values in this table, with and without the overshoot correction, are shown in Fig.\,\ref{fg:JF-disc-acc}. We include the inferred logarithmic mean DAZ accretion rate, $ \lgMdotDAZ$, and the logarithmic mean DBZ/DZ accreted mass, $ \lgMDBZ$. The number of objects in the DAZ and DBZ/DZ samples is given by $N_{\rm DAZ}$ and $N_{\rm DBZ}$, respectively. We also indicate whether the modelling included convective overshoot.
	}
	\label{tab:lifetimes}
	\begin{tabular}{lccccc}
		\hline       
		& $ \lgMdotDAZ$ & $ \lgMDBZ$ & Overshoot? & $N_{\rm DAZ}$ & $N_{\rm DBZ}$ \\
		&  [g s$^{-1}$] & [g] & [DA,DB] & & \\
		\hline
		\vspace{5pt}
		\citet{girven12} & $8.8 \pm 0.4$ & $21.9 \pm 1.1$ & $\times$ $\times$ & 13 & 8 \\
		
		\multirow{2}{*}{This work} 
		& $8.6 \pm 0.5$ & $22.4 \pm 1.4$ & $\times$ $\times$ & 19 & 11 \\
		& $8.8 \pm 0.5$ & $22.4 \pm 1.4$ & \checkmark $\times$ & 19 & 11 \\
		\hline
	\end{tabular}
	\label{tb:DiscLifetimes}\\
\end{table*}

We revisit this disc lifetime estimate using the same set of assumptions as in the \citet{girven12} study and repeat the same analysis with the now slightly larger sample of metal-polluted white dwarfs which exhibit IR excess \citep{farihi16}. {This comprises nineteen hydrogen-rich DAZ stars and eleven DBZ/DZ stars, {with the accreted masses and accretion rates shown in Fig.\,\ref{fg:JF-disc-acc}}. We consider two definitions of {the DAZ accretion rates}. One with the base of the mixed region defined at the lower Schwarzschild boundary, akin to standard 1D accretion-diffusion calculations. Secondly, we include the overshoot correction of \citetalias{cunningham19}. }

In the no-overshoot picture we find the mean accretion rate in the DAZ stars to be $\lgMdotDAZ = 8.6$ [g s$^{-1}$] and mean total accreted mass in the DBZ/DZ stars $\lgMDBZ = 22.4$ [g]. This gives rise to a disc lifetime estimate of $\log(t_{\rm disc}) = 6.3 \pm 1.4$ [yr].  We find that the new observed sample with the updated sinking times from \citet{koester14} supports a longer disc lifetime (Myr) than the earlier estimate of \citet{girven12}. Without overshoot the disc lifetime estimate has increased by almost an order of magnitude, though due to the large uncertainties in this calculation the two estimates are still consistent.

The inclusion of convective overshoot in the analysis yields a disc lifetime estimate which is just 0.2 dex shorter than its non-overshoot counterpart. This is a direct consequence of the fact that including an overshoot correction to the DAZ stars increases their mean accretion rate by 0.2 dex to $\lgMdotDAZ = 8.8$ [g s$^{-1}$]. This relatively small 3D correction is in part because a third of the DAZ stars in the sample have effective temperatures above $18\,000$\,K for which there is no overshoot correction. For white dwarfs with $T_{\rm eff}<11\,400$\,K we use the correction applied to the coolest end of the overshoot grid. {The mean total accreted mass of the DBZ/DZ stars is also expected to increase with the inclusion of convective overshoot but detailed modelling of this mixing process in helium-atmosphere white dwarfs is beyond the scope of this work.} We therefore propose a disc lifetime estimate of $\log(t_{\rm disc}) = 6.1 \pm 1.4$ [yr].

Whether or not the accretion-diffusion calculations include convective overshoot, both sets of models 
support the hypothesis of highly variable accretion rates over the Myr timescales that are typical
for metal diffusion in He-atmosphere stars.  Although the difference is now smaller with the \citetalias{cunningham19} overshoot models, the likelihood of catching a high-rate
burst in the future, in real time, may be possible with an increasingly large sample of DAZ spectra via large and ongoing surveys.  Furthermore,
these high-rate bursts of accretion should be accompanied by higher rates of X-ray production and might be detectable \citep{farihi12} by surveys such
as \textit{eRosita}.

\subsection{Predicting variability}
Identifying the parameter space of metal-polluted white dwarfs most likely to exhibit spectral variability would be of great advantage for observational campaigns. The parameter space is defined by many independent variables including, but not limited to: vertical sinking time scale, horizontal spreading time scale,  disc lifetime and accretion geometry. Given the number of independent variables, we split our discussion based on white dwarf temperature and spectral type, which are readily available observational quantities.

\subsubsection{Cool H-atmosphere white dwarfs }
Below 10\,000\,K, H-atmosphere white dwarfs have relatively large convection zones and long sinking timescales. For instance metals will stay visible for at least 10\,kyr at 7000\,K, but it could be {2--3 orders of magnitude} longer when considering the disc lifetime and phase of decreasing abundance. Fig.\,\ref{fg:homogenise05-inst.-cont.} shows that convection is efficient at spreading metals across the surface in this regime, under any accretion geometry or whether it is instaneous or continuous. Indeed, the time to homogenize the surface such that global variations are less than 50\% takes only $\sim$1000\,yr. Thus surface variations could only be observed during the {increasing}, non-steady phase of accretion, which is less than 0.1 percent of the disc lifetime. Thus the probability of finding a heterogeneous surface is very low.

\subsubsection{Warm H-atmosphere white dwarfs}
Above {13\,000\,K}, a DA white dwarf will have a convection zone which is relatively small in radius and mass ($\log M_{\rm cvz}/M_{\rm WD} \lesssim -14$) with maximum convective velocities on the order of 1\,km\,s$^{-1}$. Fig.\,\ref{fg:homogenise05-inst.-cont.} shows that such a star is unable to effectively homogenize its surface, regardless of whether accretion is instantaneous or continuous over a disc lifetime. The reason for this behaviour is that any metals accreted do not have enough time to travel across the stellar surface before they diffuse out of the convection zone.

For temperatures above 18\,000\,K, no convective instabilities are expected to develop in DA white dwarf atmospheres, i.e. those are so-called radiative atmospheres. Thus our improved convective model is not directly relevant. In this regime thermohaline instabilities may be the dominant form of {surface} mixing \citep{bauer2019}. {These instabilities have similar sizes and timescales compared to convection in the vertical direction, but it is unlikely they contribute significant spreading in the horizontal direction \citep{garaud2018}}, hence we expect spreading times in warm radiative white dwarfs to be similar to those of warm convective objects. As a consequence, it is unlikely that thermohaline mixing can lead to  significantly more homogeneous stellar surfaces than predicted in this work. 

If the accretion rate is variable in time, additional variations are expected over the sinking time, which -- in the case of white dwarfs above {13\,000\,K} -- is of the order of days to months (see Fig.\,\ref{fg:timescales}).  Radiative levitation must also be considered when predicting such temporal abundance variations.
Overall, we predict that all DAZ white dwarfs above {13\,000\,K} are expected to show surface inhomogeneity. The only known physical mechanism that would be able to prevent this outcome is near-spherical accretion geometry.

\subsubsection{H-atmospheres at intermediate temperatures}
We see in Fig.\,\ref{fg:homogenise05-inst.-cont.} that the inclusion of convective overshoot makes a considerable impact to the homogenizing efficacy of the surface layers. Because overshoot has no impact on the horizontal diffusion coefficient one can explain this as a direct consequence of the longer sinking times that result from including overshoot in the modelling (because the base of the mixed region is in deeper, slower diffusing layers).
Furthermore, solving Eq.\,\eqref{eq:2d-two-zone} for continuous accretion should in principle include the depth dependence of the horizontal diffusion coefficient (Fig.\,\ref{fg:Dderived}), which is likely to result in a relatively complex surface abundance profile with extended low abundance tails corresponding to the faster photosphere spreading. Therefore, it is difficult to predict a firm outcome for hydrogen-dominated DA white dwarfs in the range 10\,000-{13\,000\,K}.

We note that this temperature range corresponds to the quickest spreading time, with variations expected to smear out within 100\,yr (see Fig.\,\ref{fg:timescales}). This is a direct consequence of the diffusion coefficient experiencing a maximum at this temperature (i.e., the $\log g = 8.0$ track of Fig.\,\ref{fg:Dderived}), which is directly tied to this being the peak in convective velocities. However, this is somewhat balanced out by the relatively short sinking timescales on the order of 1--100 yr.

\subsubsection{Helium-rich atmospheres}
Convective instabilities in white dwarf surface layers are driven by the opacity of the dominant atmospheric constituents. For DAs the hydrogen opacity is maximal around 12\,000\,K, whereas for DBs the helium opacity peaks around 28\,000\,K. One might thus expect a similarity between the two evolutionary tracks, offset by $\approx$16\,000\,K. An examination of Fig.\,\ref{fg:ratio-DA-DB} reveals such behaviour.
We see that DB white dwarfs with effective temperatures cooler than {$\approx$30\,000\,K} are very efficient at moving accreted metals across the surface and that they are likely to be fully mixed within one diffusion timescale, similarly to cool H-atmosphere white dwarfs below 10\,000\,K as discussed above. Since very few DBZ white dwarfs are known to have temperatures above {$\approx$30\,000\,K} \citep{kepler2019}, this scenario applies for the vast majority of known DBZ and DZ white dwarfs.

\subsection{Magnetic white dwarfs}
\label{sec:mag}

\begin{figure}
	\centering
	\subfloat{\includegraphics[width=1.\columnwidth]{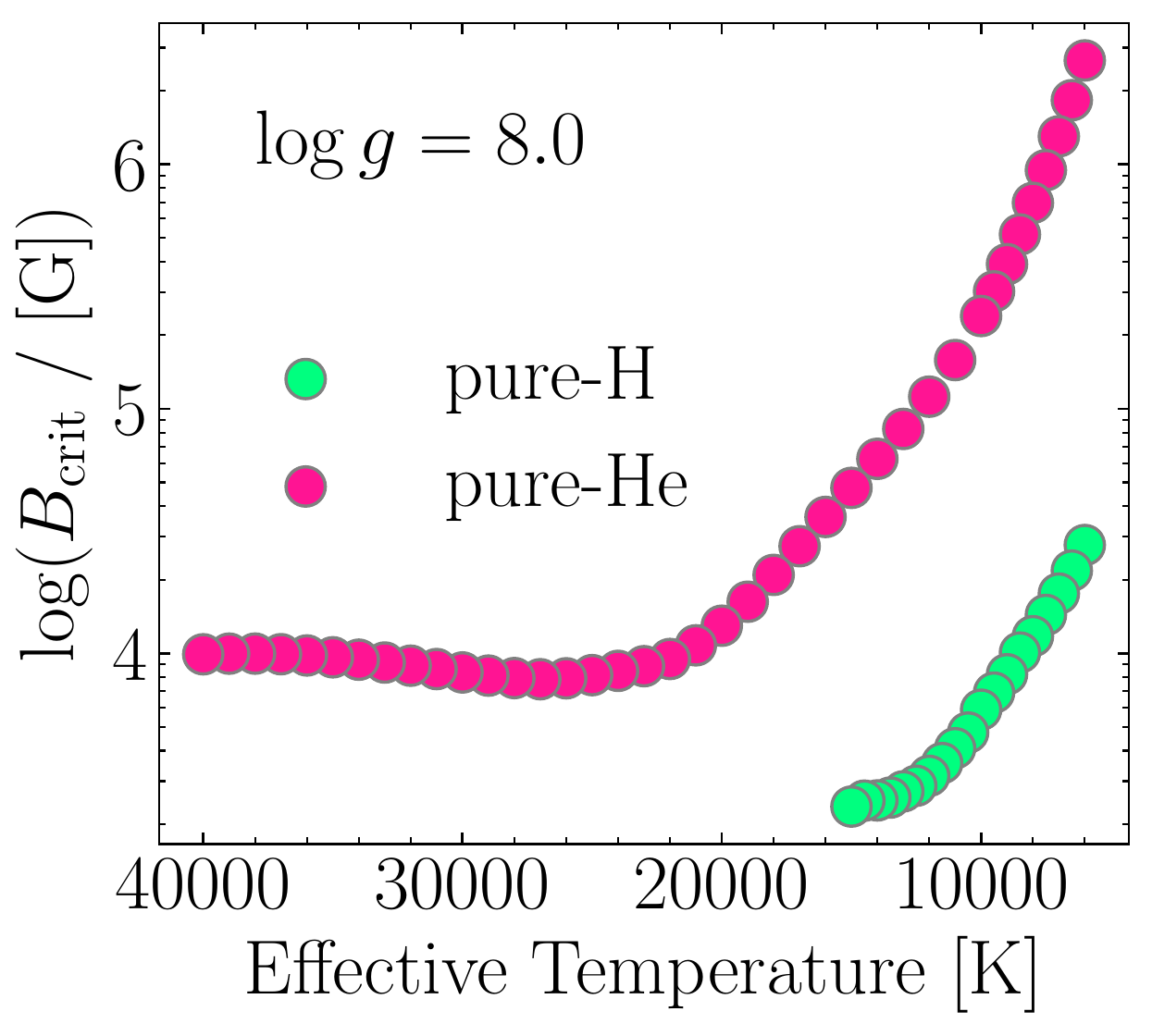}}
	\caption{{Magnetic field strength, $B_{\rm crit}$, required to produce a plasma-$\beta=1$ at the photosphere ($\tau_{\rm R}=1$) of a white dwarf as a function of effective temperature. The values are computed with Eq.\,\eqref{eq:Bcrit}, where a plasma-$\beta=1$ estimates where the magnetic fields suppress convective energy transfer. The results are shown here for pure-hydrogen (green) and pure-helium (pink) atmosphere white dwarfs with surface gravities of $\log g=8.0$. We have employed 1D model atmospheres with a mixing-length parameterization of ML2/$\alpha=0.8$ for pure-hydrogen and ML2/$\alpha=1.25$ for pure-helium composition. 	}
		}
	\label{fg:Bcrit-Tremblay15}
\end{figure}

We have discussed previously that global magnetic fields may influence the accretion of planetary material from a debris disc at metal-polluted white dwarfs. We now consider the behaviour of magnetic fields in relation to the local white dwarf atmospheric layers.
{The impact of magnetic fields on the stellar structure can be estimated from the plasma-$\beta$ parameter
\begin{equation}
	\beta = \frac{8\pi P}{B_{\rm WD}^2}\, ,
	\label{eq:plasma-beta}
\end{equation}
where $P$ is the thermal pressure and $B_{\rm WD}$ is the average magnetic field strength at the surface. {It was found from 3D magneto-hydrodynamic (MHD) simulations \citep{tremblay2015MAG} and observations \citep{gentile2018} that a value of $\beta < 1$ will inhibit convective energy transfer in the white dwarf atmosphere. Setting plasma-$\beta=1$ can therefore provide an indication of the critical magnetic field
\begin{equation}
	B_{\rm crit} = \left(8\pi P[\tau_{\rm R} = 1]\right)^{1/2} \, ,
	\label{eq:Bcrit}
\end{equation}
{\noindent}required to heavily damp convective flux in the atmosphere.
Fig.\,\ref{fg:Bcrit-Tremblay15} shows the critical magnetic field for white dwarfs with a surface gravity of $\log g=8.0$.}} Magnetic fields larger than this critical value, corresponding to 2--100\,kG in DA and DB white dwarfs, and 0.1--1\,MG in DC/DZ stars, will therefore inhibit convective energy transfer in the atmosphere. As a consequence both surface spreading and sinking of metals are likely to have different timescales. We emphasise that the lack of significant convective energy transfer does not necessarily imply that mixing is inhibited. In fact, \citet{tremblay2015MAG} demonstrated that for a pure-H atmosphere white dwarf at 10\,000\,K with an average field of 5~kG, convective velocities were of similar amplitude to the non-magnetic case even though convective energy transfer was largely suppressed. Furthermore, \alfven\ waves may contribute to additional mixing processes.
In many of the limiting cases we have discussed so far, such as cool DAZ and DZ white dwarfs, changing the diffusion coefficients by even one order of magnitude would not necessarily make surface spreading slower than sinking times, nor result in any inhomogeneity. 

Magnetic fields are also likely to influence debris disc formation and geometry of accretion. It has been shown that the \alfven\ radius, $R_{A}$, of a white dwarf with surface field strengths of $B>1$\,kG lies near or outside the typical Roche limit for accretion rates of $\dot{M}<10^{10}$\,g\,s$^{-1}$ \citep{farihi2018XRay}. In such a configuration it is highly likely that accreted gas is redirected along magnetic field lines, toward the poles. 
It was also shown that magnetic fields as weak as $B=1$\,G have an \alfven\ radius larger than the white dwarf radius for accretion rates up to $\dot{M}<10^{16}$\,g\,s$^{-1}$. There is however no observational evidence of white dwarfs with magnetic fields weaker than $\approx$ 1\,kG, or any prediction that global magnetic fields in the range 1--1000\,G could remain stable throughout white dwarf evolution including after the onset of convection.

\begin{figure}
	\centering
	\subfloat{\includegraphics[width=1.\columnwidth]{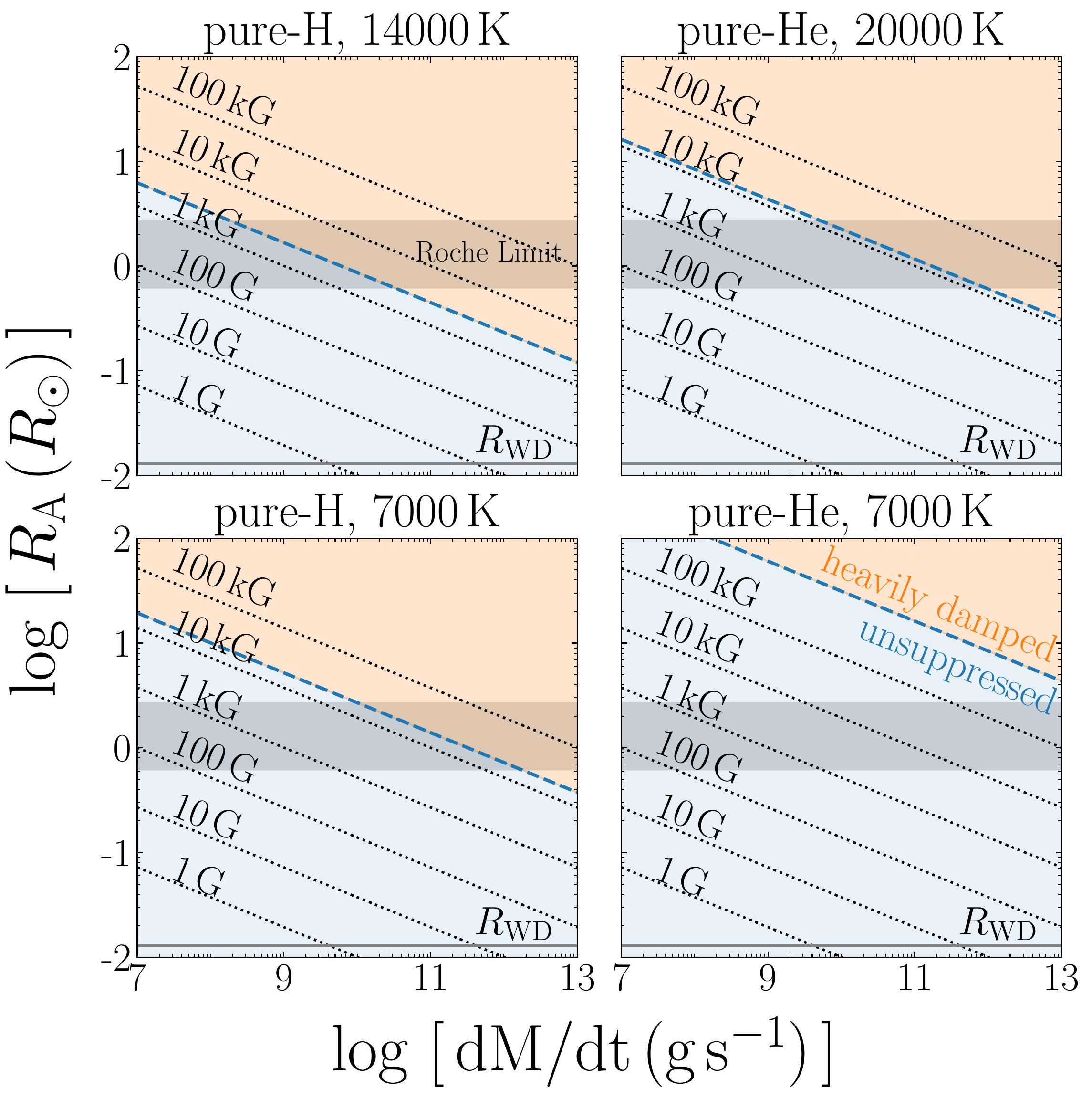}}
	\caption{{\alfven\ radius, $R_{\rm A}$, as per Eq.\,(\ref{eq:ra2}) and as function of gas accretion rate, $\dot{M}$, for various fixed strengths of dipolar surface magnetic fields (dotted lines), ranging from B=1\,G--100\,kG. The horizontal grey shaded region shows the range of Roche radii for asteroids of mean density $\rho = 1$--$8\,\mathrm{g\,cm^{-3}}$ and $C=1.3$--$2.9$, as per Eq.\,(\ref{eq:roche}). The lower horizontal grey line, near the x-axis, depicts a typical white dwarf radius of $R_{\rm WD}=0.0129\,M_{\odot}$. The magnetic field corresponding to plasma-$\beta=1$ in the photosphere is shown in dashed blue. Thus, the blue shaded region shows where magnetic fields are unlikely to impede surface convection and the orange shaded region indicates where convective flux is likely to be heavily damped by magnetic fields.
	}}
	\label{fg:Ralfven-Bcrit-Farihi18-Tremblay15}
\end{figure}

Given that magnetic fields can affect both accretion flows and atmospheric convection, we briefly consider four regimes of interest. As a reference, Fig.\,\ref{fg:Ralfven-Bcrit-Farihi18-Tremblay15}, shows \alfven\ radius, $R_{\rm A}$, as per Eq.\,(\ref{eq:ra2}), as function of gas accretion rate, $\dot{M}$, for various fixed strengths of dipolar surface magnetic fields, including the critical field $B_{\rm crit}$ as per Eq.\,(\ref{eq:Bcrit}).
\begin{enumerate}
	\item $R_{\rm A} < R_{\rm WD}$ and $B_{\rm WD}<B_{\rm crit}$ -- surface magnetic fields have no impact on accretion flows or convection. Accretion is likely to be equatorial or spherical with convective surface spreading unimpeded. This could be the case for weakly magnetic white dwarfs or systems with high accretion rates. {This is currently the default scenario for most metal-polluted white dwarfs with no measurable magnetic fields.}
	\item $R_{\rm A} > R_{\rm WD}$ and $B_{\rm WD}<B_{\rm crit}$ -- surface magnetic fields affect gas out to the \alfven\ radius, but atmospheric convection is unimpeded. Accreted gas is likely to be funnelled towards the poles and then transported via convective spreading, as described in this work, upon reaching the surface. {This scenario is plausible if weak global magnetic fields, below current observational limits, are ubiquitous in white dwarfs.}
	\item $R_{\rm A} < R_{\rm WD}$ and $B_{\rm WD}>B_{\rm crit}$ -- 
	From Fig.\,\ref{fg:Bcrit-Tremblay15}, the lowest critical magnetic field in a white dwarf is $B_{\rm crit}\approx2$\,kG. Thus, an accretion rate in excess of $\dot{M}>10^{16}$\,g\,s$^{-1}$ would be required to enable this scenario. For metal-polluted white dwarfs this scenario is therefore ruled out by observations.
	\item $R_{\rm A} > R_{\rm WD}$ and $B_{\rm WD}>B_{\rm crit}$ -- surface magnetic fields affect gas out to the \alfven\ radius and significantly dampen convective flux. Thus accretion is likely to be pole-on with convective surface spreading potentially altered, but not necessarily inhibited. \citet{landstreet2019} found that, within twenty parsec, $20\pm5$ per cent of DA white dwarfs have fields in excess of $B>30$\,kG, {which is above the critical field according to Fig.\,\ref{fg:Bcrit-Tremblay15}. Assuming this incidence does not change as a function of spectral type nor does it influence the presence of a planetary system, a significant fraction of evolved planetary systems would fall within this category.}
\end{enumerate}

\section{Discussion}
\label{sec:discussion}

One feasible way to constrain heterogeneities in the distribution of metals on a white dwarf surface is to look for photometric or spectroscopic variations on the rotational period. A spectroscopic and photometric analysis of 27 pulsating ZZ Ceti white dwarfs allowed \citet{hermes17} to constrain the rotation of single white dwarfs with masses between 0.51--0.73\,\Msun. They found that a mean period of $35 \pm 28$\,hr could explain the rotational separation of pulsation frequencies. 

Low level (few percent) photometric variations have been detected in non-pulsating white dwarfs and attributed to surface inhomogeneities that rotate in and out of view \citep{brinkworth2004,kilic15,maoz2015,hermes2017flux}. By combining \textit{Kepler} photometry with \textit{Hubble Space Telescope} (\hst) ultraviolet (UV) spectroscopy for seven white dwarfs, \citet{hallakoun18} tested the hypothesis that low level variations in the photometry could be attributed to an accretion hot spot or inhomogeneous distribution of metals across the white dwarf surface. No such correlation could be claimed although their observations did not allow to rule out the hypothesis either. In contrast, photometric variability has been linked to magnetic white dwarfs  \citep{lawrie2013,brinkworth13,reding20,gaensicke2020}. In those cases the light intensity is thought to be inhomogeneous across the stellar surface, e.g. due to an uneven temperature distribution from a yet unexplained physical mechanism related to stellar magnetism \citep{tremblay2015MAG}. As a consequence, photometric variability cannot {yet} be uniquely linked to evolved planetary systems.

The metal-rich DAZ white dwarf GD\,394 was observed to exhibit a 25 per cent modulation in its EUV flux on a period of 1.15\,d \citep{dupuis00}. The hypothesis that this variability was due to an accretion spot was tested by \citet{wilson19} with time-resolved \hst\ UV spectroscopy. They found no evidence for variations in flux, inferred accretion rate and radial velocity over the observed period. \citet{wilson2020} reported a 0.12\% flux variation in optical photometric observations from the \textit{Transiting Exoplanet Survey Satellite} (\textit{TESS}), with a period consistent with that observed by \citet{dupuis00}. They concluded that the observed variations may be explainable by a spot with an enhanced metal abundance, interaction with an orbiting planet or a magnetically-induced hot spot, but that more observations would be needed to confirm any of these three hypotheses.

The variability of accretion rates onto white dwarfs have also been studied. It was claimed that G29-38
showed evidence of a variable accretion rate based on a variation in the calcium line strength \citep{vonHippel07}. 
However, by revisiting the observations \citet{debes08} found no evidence for variation on timescales of days to years. \citet{vanderbosch2020} have recently detected marginal calcium line variability at white dwarf ZTF J013906.17+524536.89 with transiting debris. However, they conclude that the spectroscopic variability is likely related to circumstellar debris transiting along the line of sight rather than photospheric variations. All in all, there is no robust evidence of surface metal heterogeneities at white dwarfs. This is in contrast with magnetic chemically peculiar main sequence stars where inhomogeneities are routinely detected \citep[see, e.g.,][]{mobster}.

Although variations due to surface heterogeneities at white dwarfs have not yet been observed, variability of gas discs around white dwarfs is well established \citep{manser20}. In the majority of cases, the gaseous components of debris discs exhibit variations on the timescale of years to decades \citep{wilson2015,manser16,manser2016grind,dennihy2018}. Studies of dusty debris disc variability at white dwarfs have also been performed \citep{swan19,rogers20}. For a sample of white dwarfs with IR excess due to dust, \citet{rogers20} found no evidence of near-infrared $K$-band variability above the 10\% level for 32 objects with optical magnitudes smaller than 18. The upper limit on variation from the inferred dust disc was found to be 1.3\% for objects brighter than magnitude 16. In contrast, \citet{swan19} used archival data from the \textit{Wide-field Infrared Survey Explorer} (\textit{WISE}) at redder near-infrared wavelengths to show that 69\% of the 35 white dwarfs with IR excess due to dust show infrared variations more significant than 3$\sigma$. This provides important information on disc evolution and replenishment \citep{jura09,rafikov2011,metzger12,kenyon2017a,kenyon2017b}, although there is no confirmed causal connection between disc variability and accretion rate variability and geometry.

\subsection{Spectral line shapes}

For any white dwarf with heterogeneous surface metal abundances, the spectral intensity must be calculated at each point of the surface and then integrated over the stellar disc, taking into account center-to-limb variations. All regions without metals will contribute nothing to the equivalent width of a given metal line. Therefore, the maximum depth of a metal line with respect to the continuum flux is roughly proportional to the fraction of the surface covered by metals. In other words, if metals are fully confined to, e.g., 10\% of the surface, the flux decrement in any metal line should not be much larger than 10\%, with the exact maximum value depending on the region of accretion, inclination and center-to-limb coefficient. Therefore the presence of {deep or} saturated photospheric metal lines in any white dwarf spectrum {could} be sufficient to rule out any significant surface inhomogeneities. 
{We test this assumption using one example in the following section.}

\subsubsection{SDSS 1043+085}
\label{sec:geometric-syn-spectra}
{We consider in detail the expected manifestation of heterogeneity in surface metal abundances with an analysis of the  metal-polluted DAZ white dwarf SDSS J104341.53+085558.2 (hereafter SDSS 1043+085). This star has an observed IR excess due to dust and it has been shown that disc-borne gas emission lines of the Ca \I\I\ triplet are variable on a time scale of nine years \citep{gaensicke2007,brinkworth2012}.}

{For this experiment we use high-resolution UV spectroscopic observations of SDSS 1043+085. These were taken over two orbits with the \textit{HST} Cosmic Origins Spectrograph (COS; \citealt{green2012COS}) on 27 April 2015 as part of the GO program 13700 \citep{melis2017}. These observations utilised the G130M grating, with central wavelength 1291\AA, for a total exposure time of 5105\,s. More details regarding the observational setup can be found in \citet{melis2017}. The authors found evidence of C, O, Al, Si, P, S, Fe and Ni in the COS spectrum. For the purposes of our experiment, we focus only on the strongest absorption features; the Si\,\I\I\ resonance doublet at 1260 and 1264\,\AA.}

\begin{figure}
	\centering
	\subfloat{\includegraphics[width=1.\columnwidth]{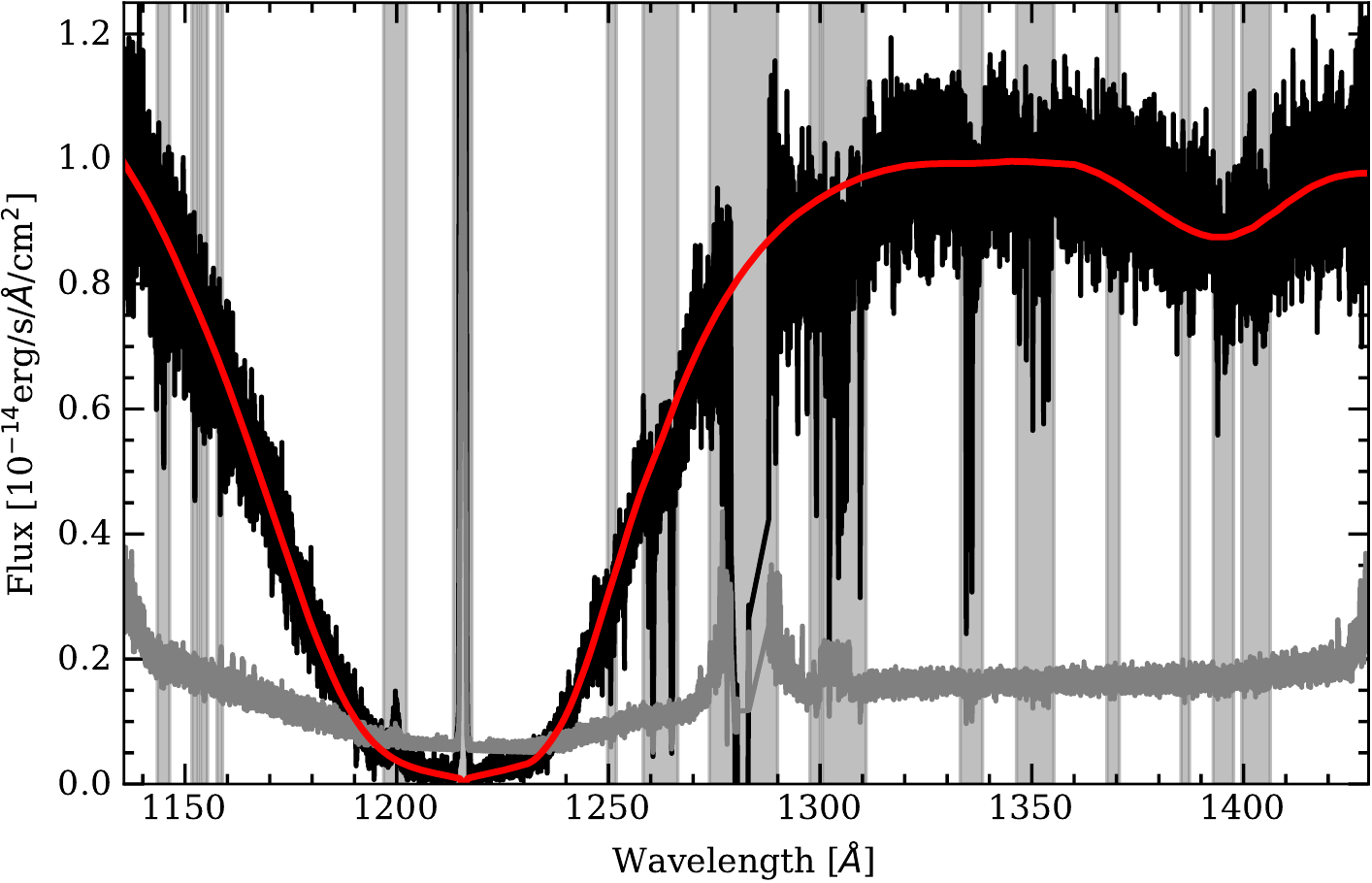}}
	\caption{{Spectroscopic observations of SDSS 1043+085 taken with \textit{HST/COS} 130M grating shown in black. The grey noisy line indicates the observational error on the flux measurement. Shown in red is the synthetic spectrum with best-fit parameters; $T_{\rm eff}=16\,536^{+76}_{-70}$\,K, $\log g=7.90^{+0.04}_{-0.03}$, $\Pi=5.98^{+0.02}_{-0.05}$~mas and reddening $E(B-V)=0.026 \pm 0.003$\,mag. This fit was made around the Ly-$\alpha$ feature, centred at 1216\,\AA, using an MCMC to find best-fit parameters for effective temperature, surface gravity, parallax and reddening. The 1-, 2- and 3-sigma tolerances on the derived parameters are shown in Fig.\,\ref{fg:SDSS1043-corner}. The vertical grey shaded regions indicate wavelengths that were masked during the fitting procedure.}}
	\label{fg:best-fit-SDSS1043}
\end{figure}

We deduce the best-fitting parameters and Si abundance for this object by performing a two-step spectroscopic fit on the \textit{HST}/COS observed spectrum shown in Fig.\,\ref{fg:best-fit-SDSS1043}. We use a grid of white dwarf models computed with the atmosphere code of \citet{koester2010} with parameters $\log g$\,=\,7.0--8.50 in steps of 0.1 dex and $T_{\rm eff}$\,=\,9000--30\,000\,K in steps of 200\,K. The grid assumes a mixing length for convection of ML2/$\alpha$\,=\,0.8 and a pure-H composition ($\mathrm{He/H} = 0$). This allows us to first update the white dwarf parameters by fitting the Ly-$\alpha$ feature centred at 1216\,\AA. The strongest metal absorption lines are excluded during this step of the fit, and the masked wavelengths are shown in shaded grey in the figure. 
We use a Markov-Chain Monte Carlo (MCMC) approach with 100 walkers that run for 5000 steps. The likelihood was set by the chi-squared value to be $-0.5\chi^2$.  We applied a Gaussian prior on the parallax based on the value from \gaia-DR2. We applied flat priors on the reddening, based on the value provided by STructuring by Inversion the Local Interstellar Medium (Stilism; \citealt{capitanio2017}) for the distance to our system. We also applied flat priors to the effective temperature and surface gravity defined by the range of the model grid.
We find the best-fitting parameters to be an effective temperature of $T_{\rm eff}=16\,536^{+76}_{-70}$\,K, surface gravity of $\log g=7.90^{+0.04}_{-0.03}$, parallax of $\Pi=5.98^{+0.02}_{-0.05}$\,mas and reddening of $E(B-V)=0.01\pm0.016$\,mag. These parameters are available in Table\,\ref{tb:SDSS1043-contiuum}.

\begin{figure*}
	\centering
	\subfloat{\includegraphics[width=1.6\columnwidth]{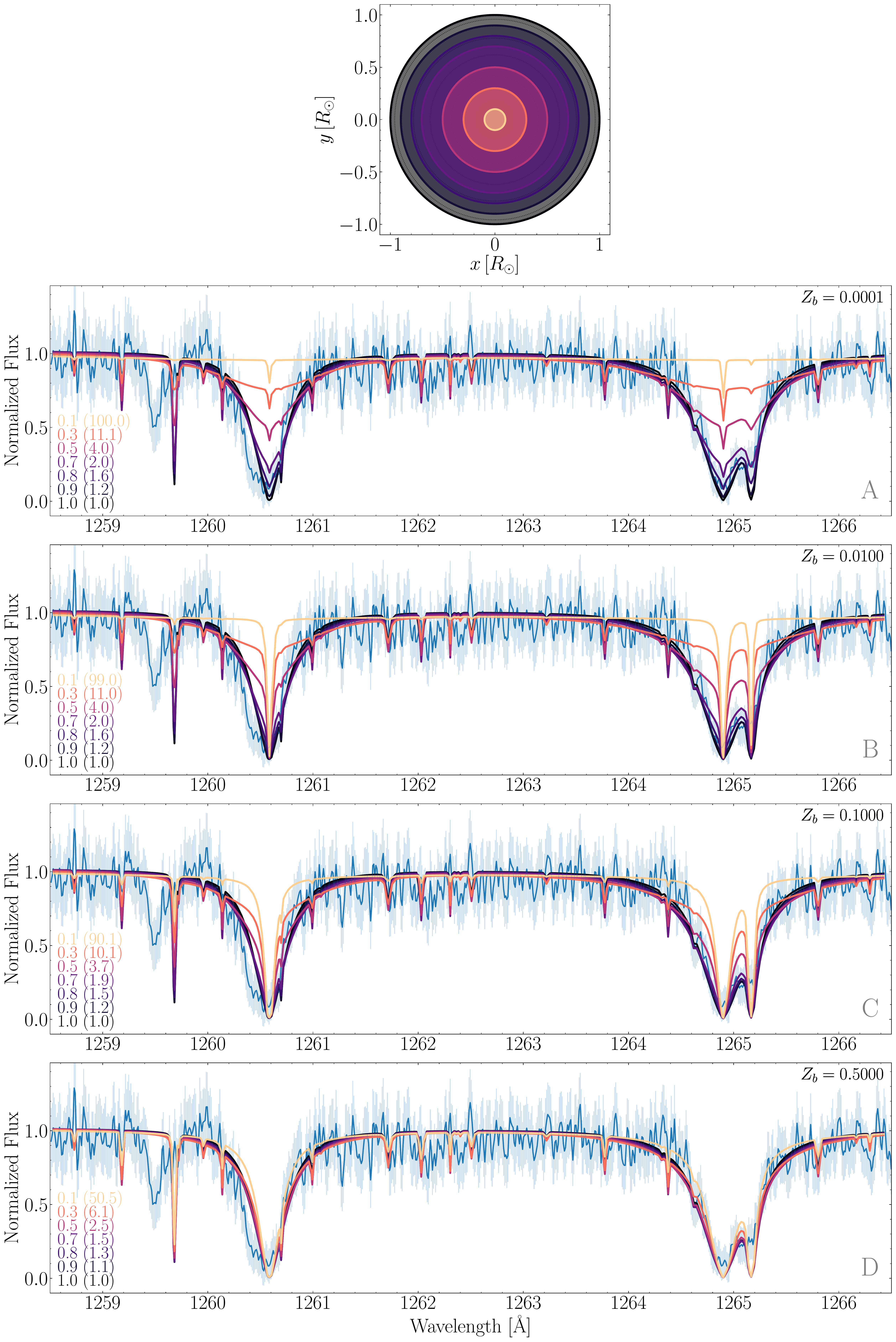}}
	\caption{
		{Spectroscopic observations of the Si\,\I\I\ resonance doublet of SDSS 1043+085 taken with \textit{HST}/COS shown in blue. 
			The black curve shows the synthetic spectrum assuming the metals are distributed homogeneously across the surface. The purple-to-yellow lines show synthetic spectra if the same amount of metals are restricted to a central spot of a given radius. The radius refers to that of a circle in the 2D projection of the stellar surface, such that the radius of the spot lies in the range $0<r\leq R_{\rm WD}$. The radii (0.1, 0.3, 0.5, 0.7, 0.8, 0.9 and 1.0 $R_{\rm WD}$) of the spots are given in the panels and correspond to surface areas of $A=$ 0.01, 0.09, 0.25, 0.49, 0.64, 0.81 and 1.00 $A_{\rm WD}$, where $A_{\rm WD}$ refers to the visible surface area of the white dwarf in the 2D plane-projection. The metal abundance of a given spot, $Z$, is scaled inversely with the area using Eq.\,\ref{eq:Zspot} and the spot metallicity fraction, $Z$, is given in brackets, next to the radius. Panels A--D vary only in the fixed background abundance of $Z_b=$0.0001, 0.01, 0.1 or 0.5. The top panel shows the spatial extent of each spot with enhanced metals, where the colours correspond to those of the synthetic spectra in the other panels. The observed spectrum (blue) has been smoothed using a Hann window function of size $N=5$, which made use of the \texttt{Python} package \texttt{PyAstronomy} \citep{pya}. The error bars shown (light blue) are the (non-smoothed) instrumental errors provided with the \textit{HST}/COS spectrum.}}
	\label{fg:geometric-spectra-SDSS1043}
\end{figure*}

The second step of our fitting procedure was to deduce the silicon abundance. Fixing the parameters derived above, an MCMC fit of the Si\,\I\I\ resonance doublet found a best-fitting abundance of $\log[\mathrm{Si/H}]=-5.27\pm0.02$. We exclude all other metal lines for the purposes of this experiment. Fig.\,\ref{fg:geometric-spectra-SDSS1043} shows the spectroscopic data from \textit{HST}/COS (blue) and the synthetic spectrum (black) with parameters adopted from Table\,\ref{tb:SDSS1043-contiuum} and the best-fitting Si abundance.

\begin{table}
	\centering
	\caption{{Best-fit parameters for SDSS 1043+085 determined by the spectroscopic fit of the Lyman-alpha feature observed by \textit{HST/COS}.}
	}
	\begin{tabular}{cccc}
		\hline            
		\vspace{2pt}
		$T_{\rm eff}$ &  $\log g$  &   $\Pi$ &  $E(B-V)$ \\
		$\mathrm{[K]}$ & [cgs] & [mas] & [mag] \\
		\hline
		$16\,536^{+76}_{-70}$  &   $7.90^{+0.04}_{-0.03}$  & $5.98^{+0.02}_{-0.05}$ &  $0.026 \pm 0.003$ \Tstrut\\
		\hline
	\end{tabular}
	\label{tb:SDSS1043-contiuum}\\
\end{table}

In order to test the hypothesis that metals may be heterogeneously distributed, we also show various synthetic spectra assuming the majority of metals are concentrated in a concentric circle on the visible stellar surface. The radius, $r$, of this circle is defined in the 2D plane projection and thus must lie in the range $R_{\rm WD} \geqslant r > 0 $. 
To conserve the total amount of metals, the metallicity of the central spot is defined as 
\begin{equation}
	Z = \frac{(1 - Z_b(1-A))}{A} \, ,
	\label{eq:Zspot}
\end{equation}
where $A=\pi r^2$ is the surface area of the spot and $Z_b$ is the background metallicity, i.e., that of the remaining stellar surface. A value of $Z_b=1$ corresponds to the original, best-fit metal abundance, and $Z_b=0$ would correspond to a zero metal abundance.

Fig.\,\ref{fg:geometric-spectra-SDSS1043} shows synthetic spectra for four background abundance values; $Z_b=0.0001,0.01,0.1\,\&\, 0.5$ in panels A--D, respectively. 
The total wavelength-dependent flux is computed by averaging the emission over the stellar surface using an eight-point Gaussian integration. 
This approach implicitly accounts for centre-to-limb darkening and should provide a reasonable estimate of the spectrum for SDSS 1043+085 under various levels of heterogeneity in its metal distribution. 
The radii of the central circles with enhanced metals 
are written in each of the lower four panels of the plot. The spot metallicity, $Z$, is also given alongside in brackets.

For a white dwarf at 16\,500\,K, our model predicts the sinking time to be at least 4 dex shorter than the lateral spreading time (see Fig.\,\ref{fg:homogenize01}).  Equations\,(\ref{eq:diff-eqn-continuous}--\ref{eq:tspread/tsink}) predict an approximate background abundance in such a case to be vanishingly small (on the order of $\sim\exp(-10^4)$). Thus, in Fig.\,\ref{fg:geometric-spectra-SDSS1043} we favour the physical parameters of Panel A (a background abundance of $Z_b=0.0001$) as being the most appropriate for this object. The lower three panels (B--D) are included as a demonstration of the possible spectroscopic appearance of a white dwarf with various metal-abundance configurations, although our model does not provide any mechanism to produce those higher background abundances for SDSS 1043+085.

Based on the synthetic spectra of Panel A, we find that, as predicted, the ratio of flux in the line cores to that in the continuum decreases with radius (and area). The background abundance contributes so little to the absorption lines that reproducing the deep flux cores is impossible, unless the spot covers most ($\gtrapprox 65$  per cent) of the surface (see the indigo line; $r=0.8$). We note that the non-linear dependence of the flux core depths on radius (or area) is a direct consequence of the centre-to-limb darkening.
We conclude that the observed metals must be fairly homogeneously distributed across the surface, a finding that somewhat contradicts the predictions of our convectively-driven lateral spreading model. This may necessitate an accretion geometry favouring a relatively homogeneous deposition pattern (i.e., close to spherical), or additional mixing processes such as differential rotation.
While we have selected a simple geometry, with the spot at the center of the disc with respect to the observer, the conclusion would be the same with a misaligned spot or an equatorial belt, with only small changes from center-to-limb effects.

We note that several other warm DAZ white dwarfs have deep or saturated metal lines in the UV. For instance, three DAZ white dwarfs observed with \textit{HST}/COS \citep{gaensicke12}, PG 0843+516, SDSS 1228+1040 and GALEX 1931+0117, have a deep Si\,\I\I\ resonance doublet at 1265\,\AA\ with near-zero flux in the line core. Based on model atmosphere fits and radial velocities, it was confirmed that the origin of these absorption lines was the white dwarf photosphere. This suggests that nearly homogeneous metal coverage may be the norm rather than the exception, even at effective temperatures in the range 19\,000--23\,000\,K, where only thermohaline mixing is expected to cause instabilities. While it is tempting to conclude that accretion must take place near-spherically in these objects, we can not rule out that we are missing another physical ingredient, e.g. another source of mixing in addition to convection and thermohaline instabilities.
 
\subsection{Constraining accretion with X-rays}

The geometry of accretion has so far received little attention in the context of the low accretion rates of metal polluted white dwarfs. 
Further work is therefore necessary to understand if the accretion process itself can be a source of mixing. Nevertheless, large enough accretion rates may lead to observational signatures such as X-ray fluxes, which is a promising method to test the accretion geometry in white dwarfs.

\citet{farihi2018XRay} presented X-ray observations of several metal-polluted white dwarfs, including WD 1145+017 and G29-38. A reanalysis of \textit{XMM-Newton} observations of G29-38
led to a new upper limit on the accretion rate based on the non-detection. Their model of X-ray production invoked a magnetically driven, pole-on accretion geometry. From optical spectropolarimetry they constrained the upper limit on the magnetic field of G29-38 to be $B=0.7 \pm 0.5$\,kG. Based on the non-detection, their X-ray accretion model found an upper limit on the accretion rate of G29-38 to be $\dot{M}=1$--$6\times10^9$\,g\,s$^{-1}$, consistent with the accretion rate of $\dot{M}=6.5 \times 10^8$\,g\,s$^{-1}$ inferred from spectroscopic abundances \citep{xu14}. The effective temperature of G29-38 is {$\approx$11\,400\,K} and the critical magnetic field for heavily damped convective flux in a DA white dwarf at this temperature is $B_{\rm crit}\approx3.5$\,kG. These parameters place G29-38 securely in scenario (ii) from Section\,\ref{sec:mag}; where $R_{\rm A} > R_{\rm WD}$ and $B_{\rm WD}<B_{\rm crit}$, implying that the lateral spreading due to convection should be well described by the results presented in this work. An examination of Fig.\,\ref{fg:homogenise05-inst.-cont.} shows that the spreading and sinking time scales are comparable for a DA white dwarf at 11\,400\,K and thus local metal abundances may be expected to vary by a factor $\sim$3 across the surface.

\begin{figure}
 \centering
 \subfloat{\includegraphics[width=1.\columnwidth]{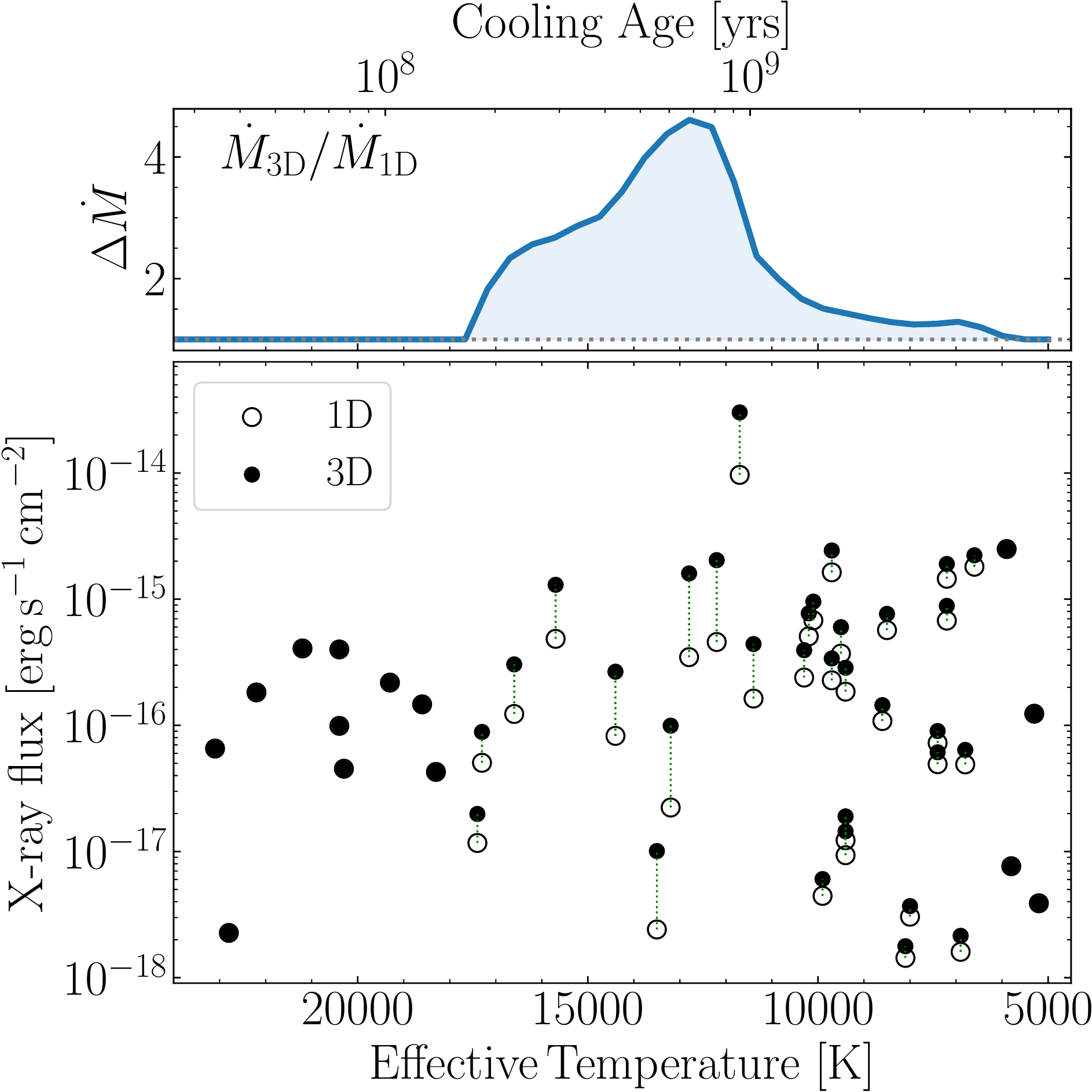}}
 \caption{{\it Bottom}: Predicted total X-ray flux reaching the Earth from 48 polluted H-atmosphere white dwarfs with published accretion rates inferred from spectroscopic observations \citep{farihi16}. The fluxes were calculated using Eq.\,\ref{eq:X-flux}, by scaling the flux upper limit for G29-38 calculated by \citet{farihi2018XRay} with the inferred accretion rate and measured parallax from \textit{Gaia}-DR2 \citep{gaia2018}. The fluxes computed using either the 1D or 3D model atmospheres \citepalias{cunningham19} and are shown in open or filled circles, respectively. The green dashed lines indicate the impact of using either of the contending models for each object with a convective atmosphere ($T_{\rm{eff}}<18\,000$ K). {\it Top}: Change in accretion rate between the 1D and 3D overshoot models. The y-axis values are $\Delta \dot{M} = \dot{M}_{\rm overshoot}/\dot{M}_{\rm 1D}$ such that no overshoot correction corresponds to $\Delta \dot{M}$ = 1 (dotted line).}
 \label{fg:xray-flux}
\end{figure}

An upper limit on the total X-ray flux at Earth was calculated, from G29-38, to be $F_{X}=9.7 \times 10^{-15}$\,erg\,cm$^{-2}$\,s$^{-1}$ \citep{farihi2018XRay}. The plasma parameters adopted for their calculation included a plasma temperature of $kT=4$\,keV and a composition equivalent to that of the Sun. Accounting for interstellar absorption and the passband of \textit{XMM-Newton} they found an upper limit on the flux to be $F_{X}=8.0 \times 10^{-15}$\,erg\,cm$^{-2}$\,s$^{-1}$. In Fig.\,\ref{fg:xray-flux} we show an estimate for the at-Earth X-ray flux from all DAZs in the \textit{Spitzer} sample. This crude estimate is found by scaling the results from \citet{farihi2018XRay} by the inferred accretion rate and \textit{Gaia}-DR2-measured parallax such that
\begin{equation}
	F_{X} \approx F_{X}^{G29} \left( \frac{\dot{M}}{\dot{M}_{G29}}\right) \left( \frac{\varpi}{\varpi_{G29}}\right)^{-2}\, ,
	\label{eq:X-flux}
\end{equation} 
where $\varpi$ is the parallax and the subscript G29 denotes the derived parameters for G29-38.

\citetalias{cunningham19} demonstrated that the inclusion of convective overshoot in the modelling of debris accretion increases the inferred accretion rate by up to an order of magnitude. A similar hypothesis was also made by \citet{bauer2019} who claimed that the inclusion of thermohaline mixing could increase the accretion rates.  
For comparison with the X-ray fluxes shown in the figure, the EPIC detector on \textit{XMM-Newton} may be sufficiently sensitive to detect a soft X-ray flux (0.5--2\,keV) of $\gtrsim$$10^{-15}\,\mathrm{erg\,s^{-1}\,cm^{-2}}$ with exposures in excess of $\gtrapprox$300\,ks (see Fig.\,3;  \citealt{watson2001XMM}). 
And according to the \textit{Chandra} Proposers' Observatory Guide\footnote{\href{https://cxc.harvard.edu/proposer/POG/html/chap6.html}{https://cxc.harvard.edu/proposer/POG/html/chap6.html}}, the point source sensitivity of the ACIS instrument on the \textit{Chandra X-ray Observatory} is $4\times10^{-15}\,\mathrm{erg\,s^{-1}\,cm^{-2}}$, for an exposure time of 104\,ks in the energy band 0.4--6.0\,keV.
Thus, current and future X-ray detecting facilities could allow the accretion rates at G29-38 and a handful of the other metal-polluted white dwarfs to be inferred via X-ray detections from sufficiently deep pointings. 
We intend to explore this line of research in a subsequent study.
The X-ray emission from accreting white dwarfs would serve as a powerful constraint on the geometry and time dependence of accretion as well as convective overshoot effects at these objects.

\section{Conclusions}
\label{sec:conclusions}

We have presented the first diffusion coefficients of stellar surface-plane transport derived from 3D tracer experiments for white dwarfs. We have utilised 3D RHD simulations for pure-hydrogen DA and pure-helium DB white dwarfs, which collectively span most of the parameter space of convective atmospheres, to examine the journey of trace metals once they are accreted to the white dwarf surface. We find a strong temperature dependence in the ability of a white dwarf to homogenize its surface composition, a finding that somewhat challenges the often employed assumption that white dwarf surface layers can be assumed to be homogeneous in chemical composition. This assumption is verified only for cooler white dwarfs, a consequence of the increasing sinking time. In H-atmosphere stars we find that surface homogeneity {can be explained by the horizontal diffusion due to convective motions} below $10\,500\pm500$\,K where the stated uncertainty takes into account the accretion geometry and time dependence. For He-atmosphere white dwarfs the surface becomes homogeneous below about 30\,000\,K.

We have demonstrated that hydrogen-rich DAZ white dwarfs above {13\,000\,K} are predicted to show metal lines only in the regions of the stellar surface 
close to where accretion occurred.
In contrast, UV observations of warm DA white dwarfs where strong and saturated metal lines are observed suggest rather homogeneous metal coverage at the surface. This discrepancy can only be explained by a near-spherical geometry of accretion or a yet unknown surface mixing process that would be several orders of magnitude faster than convective mixing. We have also provided a crude estimate of disc lifetimes around white dwarfs based on 3D RHD stellar atmosphere models, assuming a Jura-like disc and constant accretion rate, to be $\log (t_{\rm disc} / [\mathrm{yr}])=6.1\pm1.4$. This is only slightly increased compared to previous studies because the calculation depended on updated atmospheric models and a larger sample of polluted white dwarfs.

\section*{Acknowledgements}
The research leading to these results has received funding from the European Research Council under the European Union's Horizon 2020 research and innovation programme n. 677706 (WD3D). This research was supported in part by the National Science Foundation under Grant No. NSF PHY-1748958. DV  gratefully acknowledges the support of the STFC via an Ernest Rutherford Fellowship (grant ST/P003850/1). HGL acknowledges financial support by the Deutsche Forschungsgemeinschaft
(DFG, German Research Foundation) -- Project-ID 138713538 -- SFB 881 (``The
Milky Way System'', subproject A04). OT was supported by a Leverhulme Trust Research Project Grant.

\section*{Data Availability}
The observational data used in this article are published in \citet{girven12,bergfors14,farihi16,melis2017}. Outputs of previous simulation studies are published in \cite{tremblay13a,cunningham19,cukanovaite19}. Simulation data which were produced in this study will be shared upon reasonable request to the corresponding author.




\bibliographystyle{mnras}
\bibliography{mybib} 






\appendix
\clearpage
\section{SDSS 1043+085 parameter fitting}

\begin{figure}
	\centering
	\subfloat{\includegraphics[width=1.\columnwidth]{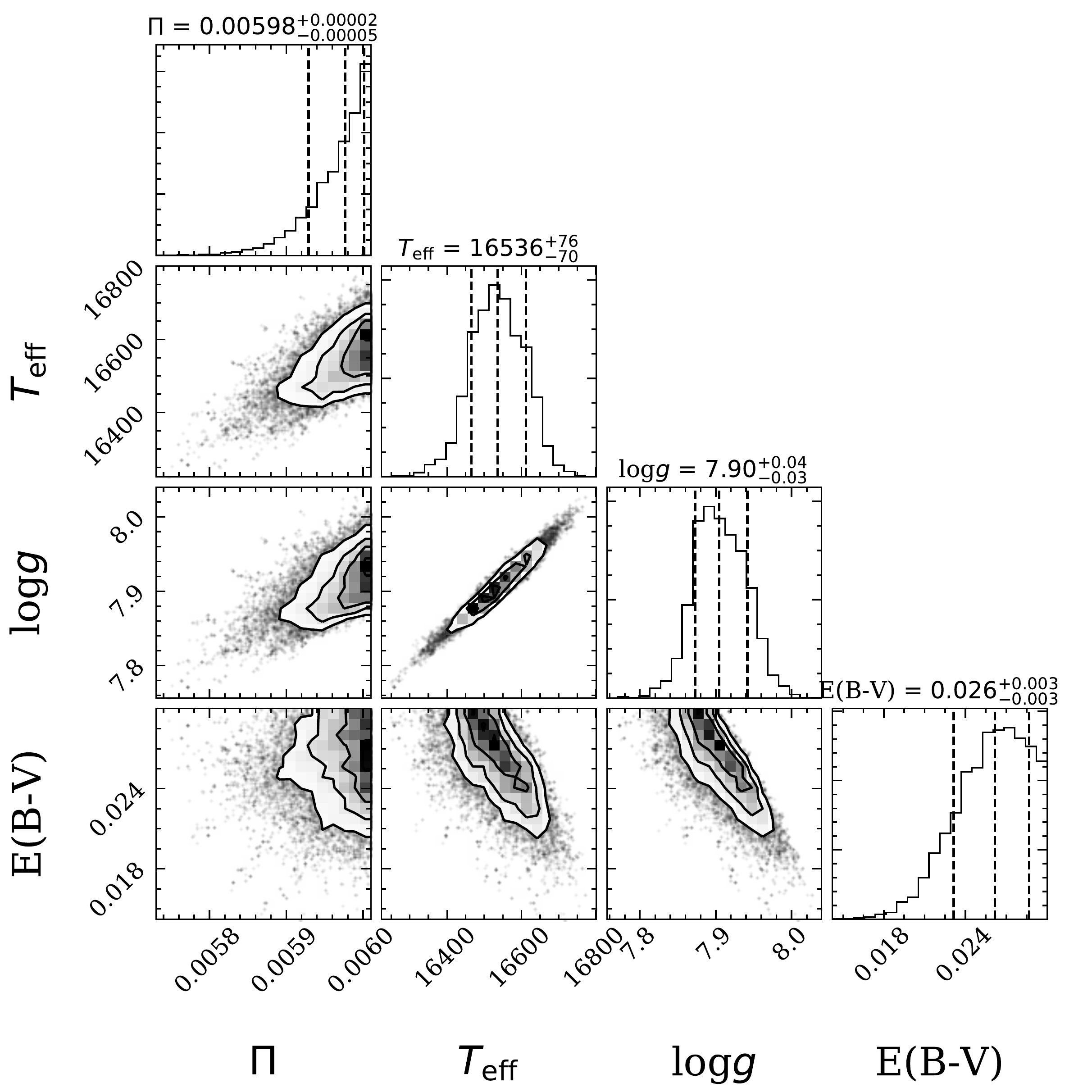}}
	\caption{\textbf{Corner plot for the Markov-Chain Monte Carlo (MCMC) leading to the best-fit parameters given in Table\,\ref{tb:SDSS1043-contiuum}. The MCMC was performed using \texttt{emcee} \citep{foreman-mackey2013} to fit the Lyman-alpha feature centered at 1216\,\AA. Strong spectral lines were masked from the fitting routine. Masked wavelengths are shown in Fig.\,\ref{fg:best-fit-SDSS1043} with grey shaded regions. This figure was made using the \texttt{corner} package \citep{corner2016}.}}
	\label{fg:SDSS1043-corner}
\end{figure}


\bsp	
\label{lastpage}
\end{document}